\documentclass[pre,twocolumn,superscriptaddress,floatfix,amssymb,showpacs,aps,10pt]{revtex4-1}
\usepackage{enumerate}
\usepackage{times}
\usepackage{graphicx}
\usepackage{amsmath}
\usepackage{color}
\usepackage{natbib}
\usepackage{array}
\newcolumntype{C}[1]{>{\centering\let\newline\\\arraybackslash\hspace{0pt}}m{#1}}

\usepackage[pdfpagemode=None,colorlinks=true,urlcolor=black,%
linkcolor=blue,citecolor=blue,pdfstartview=FitH]{hyperref}
\bibpunct{[}{]}{,}{n}{}{;}
\usepackage{lipsum}
\usepackage{bm}
\usepackage{multirow}

\newcommand*{\vel}{\ensuremath v}
\newcommand*{\solvpos}{\mathbf x}

\newcommand*{\beadpos}{\solvpos}

\newcommand*{\potential}{\ensuremath V}

\newcommand*{\ljsigma}{\ensuremath \sigma}

\newcommand*{\fenepot}{\potential_\mathrm{FENE}}

\newcommand*{\Rg}{\ensuremath{R_g}}

\newcommand*{\Npoly}{\ensuremath N}

\newcommand*{\flory}{\ensuremath \nu}

\begin{document}


\title{Driven polymer translocation in good and bad solvent: effects of hydrodynamics and tension propagation}


\author{J. E. Moisio}
\affiliation{GE Healthcare, Kuortaneenkatu 2, FI-00510 Helsinki, Finland}
\author{J. Piili}
\author{R. P. Linna}
\email{Author to whom correspondence should be addressed: riku.linna@aalto.fi}
\affiliation{Department of Computer Science, Aalto University, P.O. Box 15400, FI-00076 Aalto, Finland}



\date{\today}

\begin{abstract}
We investigate the driven polymer translocation through a nanometer-scale pore in the presence and absence of hydrodynamics both in good and bad solvent. We present our results on tension propagating along the polymer segment on the {\it cis} side that is measured for the first time using our method that works also in the presence of hydrodynamics.  For simulations we use stochastic rotation dynamics, also called multi-particle collision dynamics. We find that in the good solvent the tension propagates very similarly whether hydrodynamics is included or not. Only the tensed segment is by a constant factor shorter in the presence of hydrodynamics. The shorter tensed segment and the hydrodynamic interactions contribute to a smaller friction for the translocating polymer when hydrodynamics is included, which shows as smaller waiting times and a smaller exponent in the scaling of the translocation time with the polymer length. In the bad solvent hydrodynamics has a minimal effect on polymer translocation in contrast to the good solvent where it speeds up translocation. We find that under bad-solvent conditions tension does not spread appreciably along the polymer. Consequently, translocation time does not scale with the polymer length. By measuring the effective friction in a setup where a polymer in free solvent is pulled by a constant force at the end, we find that hydrodynamics does speed up collective polymer motion in the bad solvent even more effectively than in the good solvent. However, hydrodynamics has a negligible effect on the motion of individual monomers within the highly correlated globular conformation on the {\it cis}-side and hence on the entire driven translocation under bad-solvent conditions.
\end{abstract}


\maketitle

\section{\label{sec:introduction}Introduction}

The process of a polymer driven through a nanometer-scale pore by a force acting inside the pore is well understood under good-solvent conditions. It is well established that the polymer is driven out of equilibrium even for moderate pore force. On the {\it cis} side, that is, on the side from where the polymer translocates, this shows as the tension propagating along the polymer chain~\cite{PhysRevE.76.021803,lehtola09,PhysRevE.78.061803,PhysRevE.85.051803,suhonen14} and on the {\it trans} side, {\it i.e.} the side where the polymer translocates to, as crowding of polymer segments~\cite{lehtola09}. We have previously shown that this crowding has no discernible effect on the driven translocation~\cite{suhonen14}. The scaling of the translocation time with the polymer length $\tau \sim N^\beta$ in the good solvent follows from the scaling of the length to which the tension has propagated $n_t$ with the number of translocated monomers, or beads, $n_t \sim s^{\beta-1}$. If $\tau$ scaled with $N$ also under bad-solvent conditions, tension propagation would be the most likely explaining mechanism. 

Hydrodynamics has been shown to speed up driven polymer translocation under good-solvent-conditions~\cite{fyta06:_multis,melchionna07:_explor_dna_lattic_boltz,lehtola09,PhysRevE.78.061803}. The significance of hydrodynamic interactions under bad-solvent conditions has not been established. As will be seen hydrodynamic interactions is a good way of characterizing how the motion of the polymer segment on the {\it cis} side takes place.

Previously, we have measured tension propagation along the polymer chain alternatively via the motion of the polymer beads~\cite{lehtola09} or the strain of individual bonds~\cite{suhonen14}. Except for the preliminary results in~\cite{lehtola09} tension propagation in the presence of hydrodynamics has not been studied. The first-mentioned indirect measurement is the only method that has been used in the presence of hydrodynamics. However, since in the presence of hydrodynamics monomers may be set in motion toward the pore before the tension has reached them, this method does not give the true true dynamics of the propagating tension when hydrodynamics is involved. Encouraged by the very precise measurements of the tension using the second, more direct method in the absence of hydrodynamics, we will apply it here also in the presence of hydrodynamics. We will determine the tension spreading dynamics and also how fast hydrodynamic interactions set in compared with the tension propagation speed in the good solvent.

The paper is organized as follows: The computational model is explained in Section~\ref{sec:model}. First, the implementation of the polymer and solvent dynamics is covered in subsection A. The polymer model, the indirect implementation of the solvent quality using this polymer model, and the simulation geometry are explained in subsections B, C, and D, respectively. Results are reported and analyzed in Section~\ref{sec:results}. Subsections A and B cover measurements of translocation time and radii of gyration, respectively. These first subsections set the background and open questions for our research. The main findings are covered in subsection C. Here our results from measurements of waiting times, tension, and friction are reported and analyzed. This subsection is further divided in two subsections that cover the results pertinent to the good and bad solvent separately.

\section{\label{sec:model} The Computational Model}

\subsection{\label{sec:srd}Polymer and solvent dynamics}

The dynamics of the polymer immersed in the solvent is implemented by a hybrid method where the polymer beads perform molecular dynamics (MD), whose timestep $\delta t = 0.001$, and at every $1000$ MD timesteps both the polymer and the solvent perform stochastic rotation dynamics (SRD), whose timestep $\Delta t = 1$. The timesteps along with the distance and potential values that follow are given in reduced units, see {\it e.g.} \cite{allen}. The polymer's equations of motion are integrated in time by the velocity Verlet algorithm~\cite{swope_velocity_verlet,FrenkelSmit}.

SRD that is used to implement solvent dynamics supports hydrodynamic modes~\citep{JChemPhys.110.8605}. In the SRD model the simulation space is divided into cubic lattice of cells, whose sides are of unit length. Each cell contains on average $5$ solvent particles (SP) and one polymer bead (PB). SPs are fictitious in the sense that they can be thought of as carrying the mass and momenta of multiple realistic particles. The interactions of SPs and PBs are approximated in a stochastic fashion as described below.

An SRD step consists of streaming and collision steps. In the streaming step the position of each solvent particle is propagated according to
\begin{equation}
\mathbf{x}_i{(t + \Delta t)} = \mathbf{x}_i(t) + \mathbf{v}_i(t) \Delta t.
\end{equation}
After this the collisions between the solvent particles and the polymer beads are taken into account by 
\begin{equation}
\mathbf{v}_i{(t + \Delta t)} = \mathbf{v}_{\rm COM}(t) + \mathbf{\Omega} \left( \mathbf{v}_i(t) - \mathbf{v}_{\rm COM}(t) \right) ,
\end{equation}
where $\mathbf{v}_{\rm COM}$ is the velocity of the center of mass of the solvent particles and polymer beads within the cell they currently belong to and $\mathbf{\Omega}$ is a stochastic rotation matrix. The rotation angle is constant but the direction of the axis with respect to which the velocities are rotated at each timestep is chosen randomly for each cell. The lattice of cells has periodic boundary conditions.

To maintain molecular chaos, that is, in order to avoid artificial correlations between solvent particles, random grid shifts were applied to the cells before the collision step~\citep{PhysRevE.63.020201,PhysRevE.67.066705}. In the grid shift the cells (or equivalently the particles) are moved by a displacement between $[-0.5, 0.5]$ of the cell dimension sampled from uniformly random distribution.

A particular advantage of the SRD method is that it allows the switching off of hydrodynamic interactions by randomly shuffling the momenta of the solvent particles after each collision step. This feature is crucial in determining the effects of hydrodynamics on which it is very hard to obtain quantitative information. To pin down the effects of hydrodynamic interactions we will make close comparison between translocation dynamics taking place in a Brownian heat bath and in the presence of full hydrodynamic interactions.

\subsection{\label{sec:polymodel}Polymer model}

In order to gain understanding on the dynamics of polymer translocation under bad solvent conditions we choose a flexible polymer model. The used freely-jointed chain (FJC) model is commonly used to describe single-stranded DNA, RNA, and proteins. 

Forces for the equations of motion integrated in time using the velocity Verlet method are obtained from the potentials for the interconnected point-like beads in FJC as $\mathbf{f} = -\nabla V$. Consequent beads in FJC are connected by the anharmonic FENE potential given by
\begin{equation}
\fenepot = -\frac{H}{2} R_0^2  \ln \left( 1 -
\frac{r^2}{R_0^2} \right) ,
\end{equation}
where $H$ is a parameter describing the strength of the potential, $r$ is the distance, and  $R_0$ is the maximum distance between consecutive beads allowed by the FENE constraint. Lennard-Jones potential acts between all PBs
\begin{align}
U_{\rm LJ} = \left\lbrace \begin{array}{lcl}
         4 \epsilon \left[\left(\frac{\sigma}{r_{ij}}\right)^{12} -
\left(\frac{\sigma}{r_{ij}}\right)^6\ \right] + \epsilon(1-Q) &,& r_{ij}
\leq  r_c\\
         4 \epsilon \left[\left(\frac{\sigma}{r_{ij}}\right)^{12} -
\left(\frac{\sigma}{r_{ij}}\right)^6\ \right]Q        &,& r_{ij} > r_c
         \end{array}
  \right.,
\label{eq:ljpot}
\end{align}
where $\epsilon = 1.2$ is a coupling constant, $\sigma = 1$ sets the length scale of the interactions, $Q$ sets the solvent quality, and $r_c$ is the cut-off distance for the interaction in the case of good solvent (see the next subsection). 

\subsection{\label{sec:solvent}Implementation of solvent quality}

The solubility of the polymer determines the initial polymer conformation and, as we will see, has a strong effect on the translocation dynamics. In a good solvent a polymer spreads out and the polymer contour forms a self-avoiding random walk. In contrast, bad solvent is repelled from within the globular polymer conformation. In SRD, due to the solvent being implemented by fictitious particles that represent the explicit solvent particles the solvent quality has to be defined indirectly with the aid of monomer interactions. More specifically, LJ potential is used as an effective substitute for real hydrophilic or hydrophobic interactions, since the SRD model for fluid dynamics does not incorporate the complex electrostatic interactions between real solvent and polymer molecules, in contrast to, for example, dissipative particle dynamics (DPD)~\cite{yang13}.

A polymer in good solvent can be simulated by excluding the attractive interactions between monomers. This is done in a standard way by truncating the LJ potential at $r_c = 2^\frac{1}{6} \ljsigma$, where $V_{LJ} = 0$, which means setting $Q = 0$ in Eq.~(\ref{eq:ljpot}). For a polymer immersed in bad solvent the full form of Eq.~(\ref{eq:ljpot}) that includes both the repulsive and the attractive parts of the potential is used, {\it i.e.} $Q = 1$. The attraction between monomers corresponds to the repulsion between the polymer and the solvent leading to a globular polymer conformation (see the first snapshot on the second row in Fig.~\ref{fig:snapshots}). While this indirect implementation has its limitations, causing unphysical artefacts when solvents of two different qualities are used in the same simulation \citep{kapahnke10:_polym}, it is valid in any simulation where there is solvent of only one quality present at any time, which is the case in the present study.

\subsection{\label{sec:transgeom}Translocation geometry and pore model}

\begin{figure}
\includegraphics[width=0.5\textwidth]{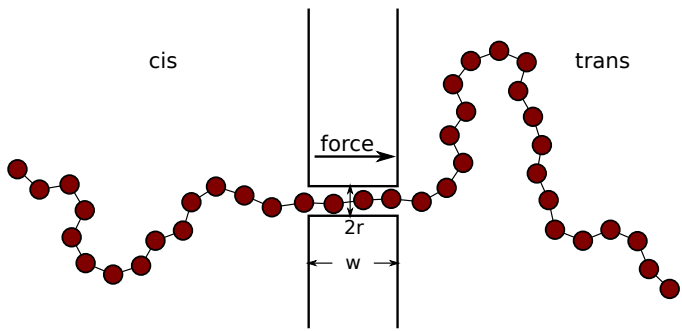}
\caption{\label{fig:transgeometry}Schematic depiction of the geometry used in the translocation simulations.}
\end{figure}

The translocation geometry is depicted in Fig.~\ref{fig:transgeometry}. The polymer is initially on the {\it cis} side with three beads inside the pore. An infinite wall parallel to the $xy$ plane divides the simulation space. No-slip boundary conditions bounce the solvent particles and the polymer beads from the wall making it impenetrable for them. The polymer can pass from the {\it cis} to the {\it trans} side only through a cylindrical pore, whose axis is parallel to the $z$ axis. 

While there is no solvent in the pore, every bead inside the pore experiences a damping force $\mathbf{F}_\gamma = - \gamma \mathbf{v}$, where the damping constant $\gamma$ is chosen such that the damping inside the pore corresponds to the damping caused by the solvent according to the damping coefficient calculations by~\citet{kikuchi03:_trans}. A harmonic pore force $\mathbf{F}_h = -k \mathbf{l}$ applied to each bead inside the pore keeps the polymer aligned and prevents hair pinning. Here $\mathbf{l}$ is the displacement vector from pore axis to the bead (perpendicular to the pore axis) and $k = 100$. The aligning force increases slightly the effective driving pore force. We have checked that changing $k$ from $100$ to $1000$ has no appreciable effect on translocation characteristics. The aligning force is introduced primarily for numerical reasons. Its physical origin would be the repulsive potential of the pore surfaces. Previously, we have verified that the harmonic pore and the commonly used bead pore give identical characteristics for the driven polymer translocation~\cite{PhysRevE.78.061803}.

The translocating polymer is driven by force exerted on the polymer beads inside the pore. The reported driving forces denote the forces applied to each bead
inside the pore. Hence, the force applied to the whole polymer is the reported
force multiplied by the number of beads inside the pore, which on average is
$w / \sigma \approx 3$, where $w$ is the thickness of the wall and $\sigma$ is the mean distance of the beads. The driving forces considered in this work are in the moderate to large regime ($F \sigma / k_BT \ge 1$) according to the definition by~\citet{PhysRevE.85.041801}. 


\section{\label{sec:results}Results}

\begin{figure}
\includegraphics[width=0.13\textwidth]{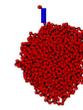}
\caption{\label{fig:snapshots}(Color online) {\bf In this arXive version only the initial conformation in bad solvent is shown due to size limitations.} Snapshots of translocating polymers of length $N=1600$. Initial conformations on the left. Pore force $F = 1$. Hydrodynamics is included. {\bf The first row:} good solvent. {\bf The second row:} bad solvent. The polymer translocates from the bottom {\it cis} side to the top {\it trans} side. On the second row the pore region is shown by a blue rectangle. The wall whose thickness is equal to the pore length extends horizontally to the left and right from the pore (not shown). Due to a different length scale the pore does not show on the first row. In the leftmost snapshot only three monomers are inside the pore on the top. In the following snapshots the pore and the wall lie just below the upper region of crowded monomers.}
\end{figure}

In what follows the focus is in the detailed measurements of tension and waiting times that are presented in subsection~\ref{sec:tension}. The first subsections~\ref{sec:tauscaling} and \ref{sec:rg} set the background and questions for our study. It is seen from Fig.~\ref{fig:snapshots} showing series of snapshots of the simulated driven translocations in the good and bad solvent how different the process is depending on the solvent quality.

\subsection{\label{sec:tauscaling}Translocation time}

Polymers of lengths $N = 25, 50, 100, 200, 400, 800, 1600$ beads were driven through the pore by a force $F$ acting inside the pore. The amount of computation needed for equilibrating the longest chains $N = 800$ and $1600$ proved to be extensive. Therefore, the simulations for these polymer lengths were run in parallel using multiple CPU cores. The results were obtained by averaging over at least $200$ translocations for chains of length $N \le 400$ and at least $50$ translocations for $N \ge 800$.


\begin{figure}
\includegraphics[width=0.23\textwidth]{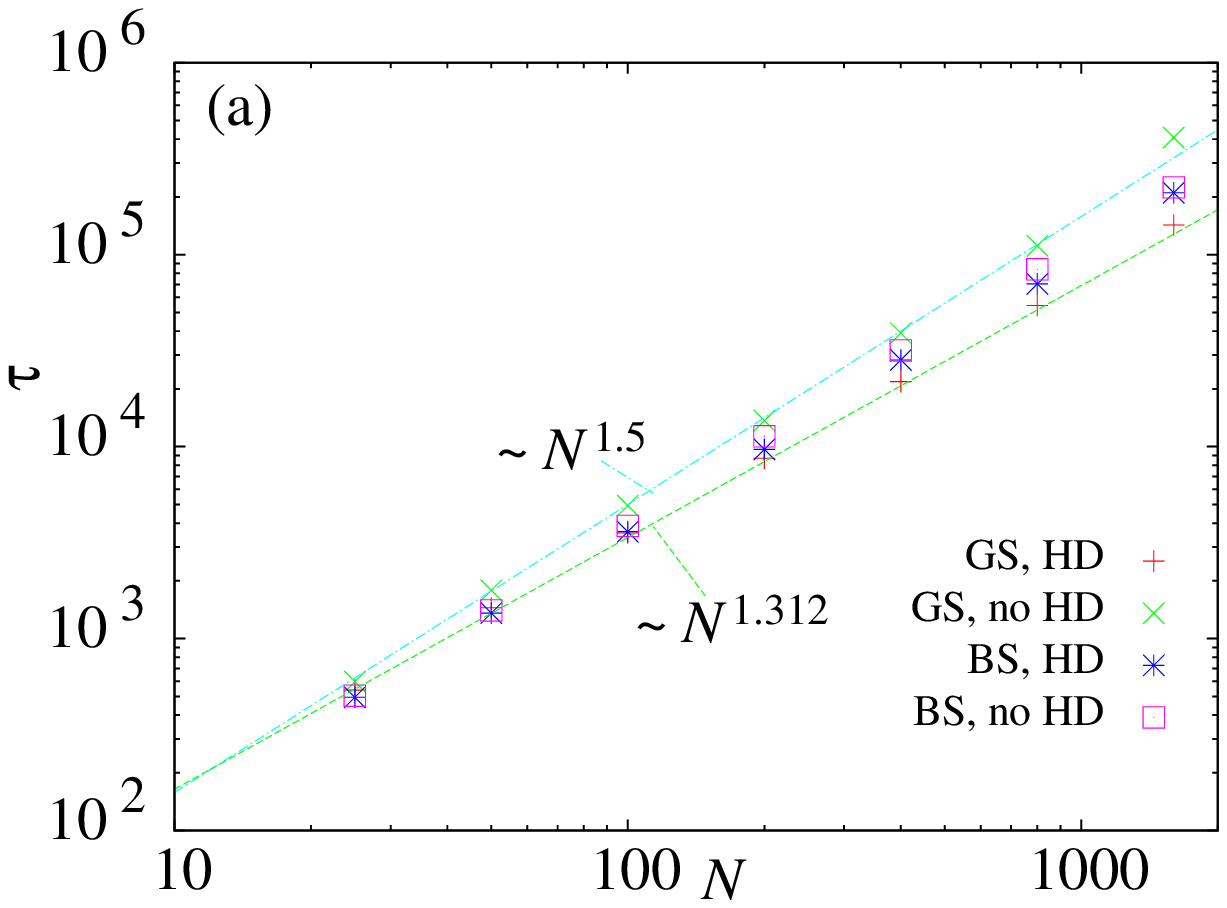}
\includegraphics[width=0.23\textwidth]{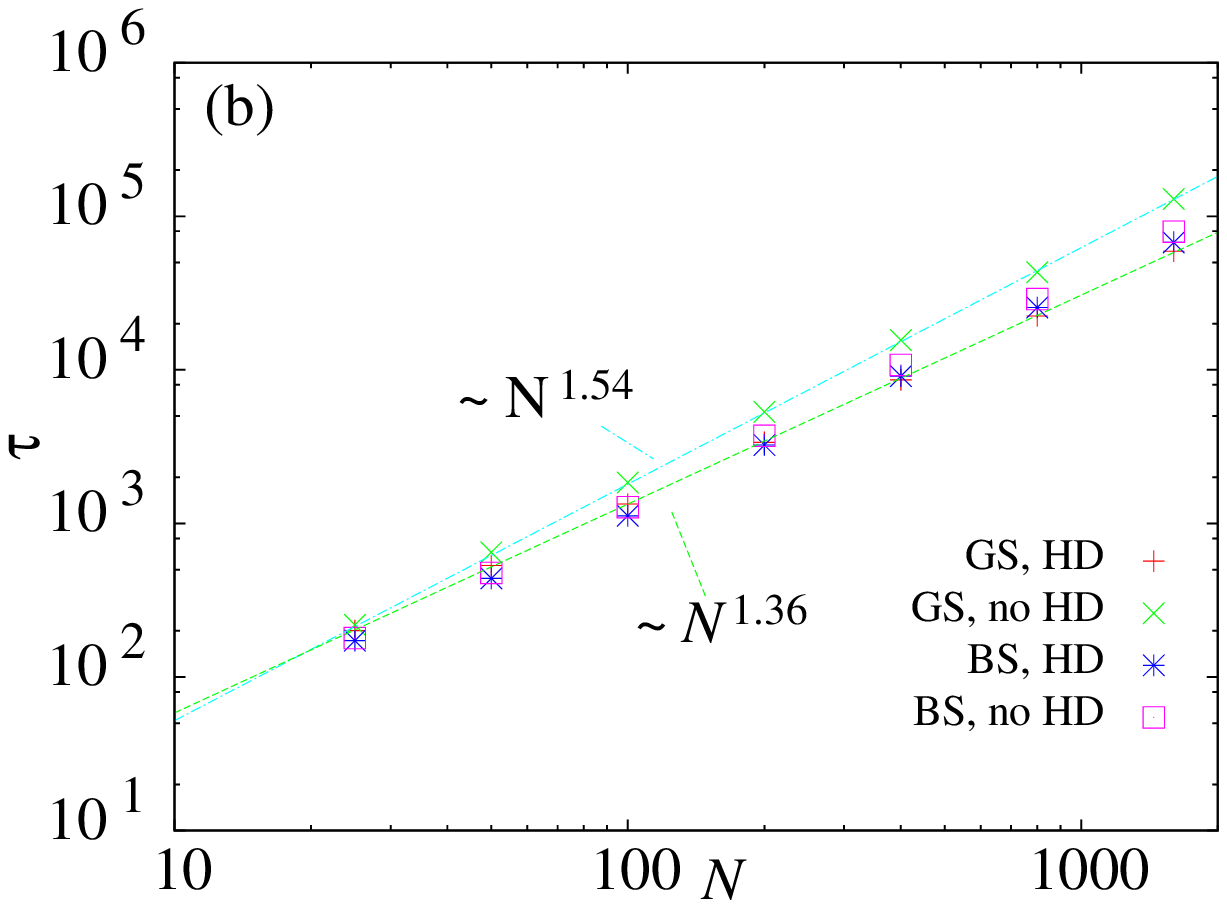}
\caption{\label{fig:tau}(Color online) Translocation times $\tau$ as a function of polymer length $N$ for good (GS) and bad solvent (BS) with and without hydrodynamics (HD). (a) $F = 1$. (b) $F = 3$.}
\end{figure}

As expected, clear scaling of translocation time with the polymer length, $\tau \sim N^\beta$, was obtained for the good solvent case, see Fig.~\ref{fig:tau}. For the large pore force of $F = 10$ stalling events due to local jamming at the pore entrance deteriorate the perfect scaling in the good solvent (not shown). In the good solvent we obtain $\beta \approx 1.5$ and $1.54$ for $F = 1$, $3$, respectively in the absence of hydrodynamics. The  measurements of $\beta$ for $F = 1$ and $3$ not affected by jammings are in accordance with the many times confirmed increase of $\beta$ with $F$~\cite{lehtola09,suhonen14}. We showed that this increase of $\beta$ with $F$ comes from the {\it trans} side and addressed it to fluctuations that were shown by Dubbeldam {\it et al.} to assist translocation~\cite{dubbeldam13}. Hydrodynamic interactions decrease $\beta$ as we have found previously~\cite{lehtola09}. We obtain $\beta \approx 1.31$ and $1.36$ for $F = 1$ and $3$, respectively, when hydrodynamics is included.


In the case of bad solvent the situation is not as clear. $\tau$ depends on $N$ in a way where a mechanism resulting in $\tau$ scaling with $N$ may play a role, see Fig.~\ref{fig:tau}. However, this mechanism, if there is one, does not dominate the overall translocation dynamics. In simulations made using DPD~\cite{yang13} scaling $\tau \sim N^\beta$ was not obtained for polymer translocation in bad solvent. We will look into this in more detail in the following sections.

In accordance with previous findings~\cite{fyta06:_multis,lehtola09} hydrodynamic interactions are seen to speed up translocation in good solvent, see Fig.~\ref{fig:tau}. Hydrodynamics does speed up translocation also under bad solvent conditions. However, here the speed-up is small. In the bad solvent where the polymer is in globular conformation tension does not appreciably propagate, as will be seen in Section~\ref{sec:tension}. The increased correlations due to hydrodynamics that enhance momentum transfer along the extended polymer chain in good solvent may not contribute appreciably to the motion of monomers screened by the globular polymer conformation in bad solvent. In addition, it is an open question how effectively hydrodynamics enhances motion in the bad solvent in general. We will determine these issues in Section~\ref{sec:bs}.

Regardless of the solvent quality hydrodynamics speeds up the translocation more effectively for large $F$. The momentum that is mediated along the polymer chain from the pore to the {\it cis} side is larger for larger $F$. Drift due to bias dominates over diffusion more strongly for large $F$. Hence, it is clear that  the effect of hydrodynamics via the mediated momentum is more pronounced for large $F$. The reduction of friction due to hydrodynamics will be dealt with in Section~\ref{sec:tension}.

\subsection{\label{sec:rg}Radius of gyration}

Radius of gyration defined as
\begin{equation}
\Rg = \sqrt{\frac{1}{N} \sum_{i=1}^N
\left(\mathbf{x}_i - \mathbf{x}_{\textrm{COM}} \right)^2},  \label{eq:rg}
\end{equation}
where $\mathbf{x}_i$ and and $\mathbf{x}_{\textrm{COM}}$ are the positions of bead $i$ and the center of mass of the polymer, respectively, can be used to determine if a polymer is driven out of equilibrium during translocation~\cite{lehtola09,bhattacharya10,suhonen14}. For polymers in equilibrium the relation $R_g^{eq} \propto N^\nu$ holds, where $\nu$ is the Flory exponent. We measured $\nu \approx 0.63$ and $\nu \approx 0.31$ for equilibrated polymer conformations in good and bad solvent, respectively.

\begin{figure}
\includegraphics[width=0.23\textwidth]{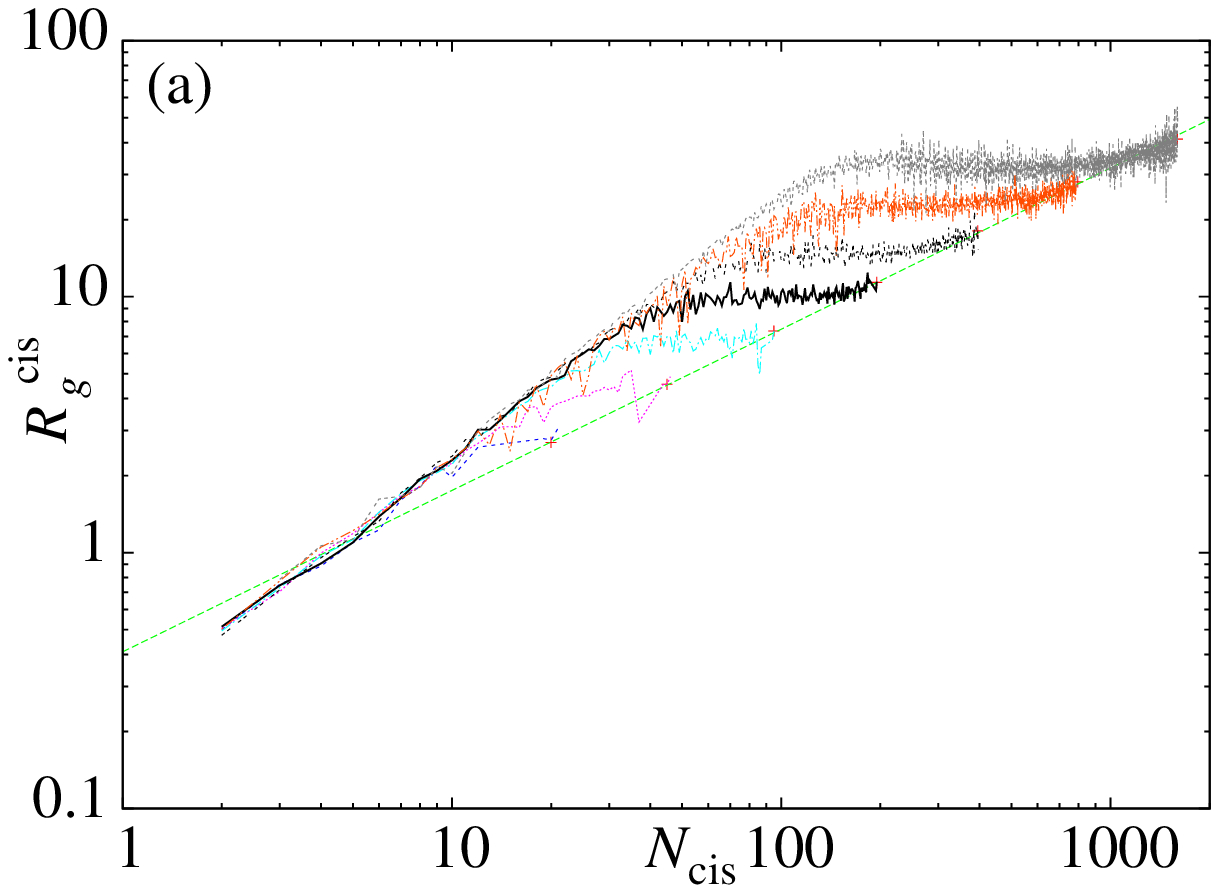}
\includegraphics[width=0.23\textwidth]{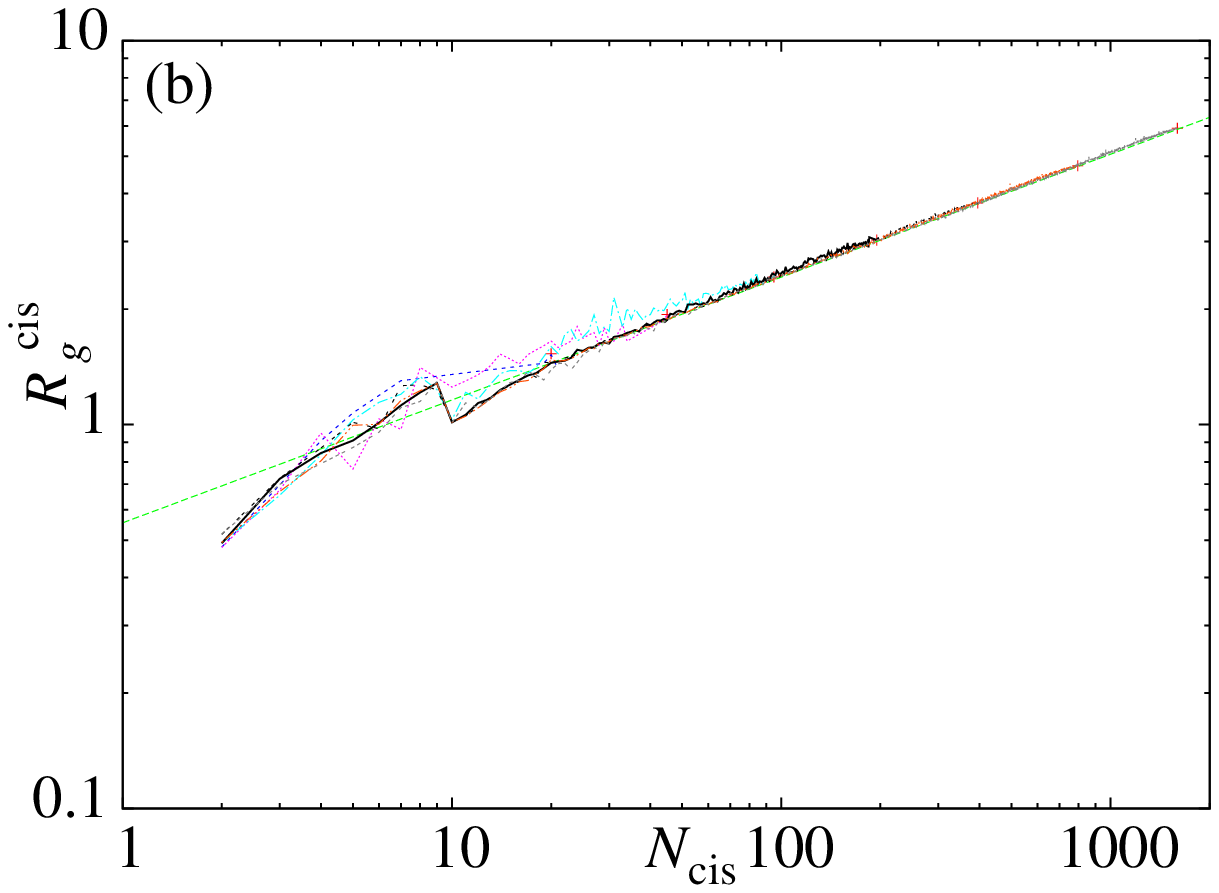}
\caption{\label{fig:transrg}Measured radii of gyration of the translocating polymers on the {\it cis} side. The equilibrium $R_g$ measured for seven discrete polymer lengths and the solid line giving the obtained scaling $R_g \sim N^{\nu({\rm measured})}$ are shown for reference. Hydrodynamics is included. $F = 10$. (a) Good solvent, $\nu({\rm measured}) \approx 0.63$. (b) Bad solvent, $\nu({\rm measured}) \approx 0.32$. The chain lengths are (from bottom to top in (a)) \mbox{$\Npoly = 25, \; 50, \; 100, \; 200, \; 400, \; 800, \; 1600$.}
}
\end{figure}

Fig.~\ref{fig:transrg} shows $R_g$'s measured during the translocations as a function of the number of beads on the {\it cis} side $N_{cis} = N - s$ for polymer conformations when hydrodynamics is included. As expected, $R_g$ of the translocating polymers in good solvent show significant deviation from the equilibrium scaling. The polymers in bad solvent maintain their globular equilibrium conformation throughout the translocation process. Small deviation of $R_g$ from the equilibrium scaling can be seen at the end (for small $N_{\rm cis}$) when the tail of the polymer is sucked into the pore.

\begin{figure}
\includegraphics[width=0.23\textwidth]{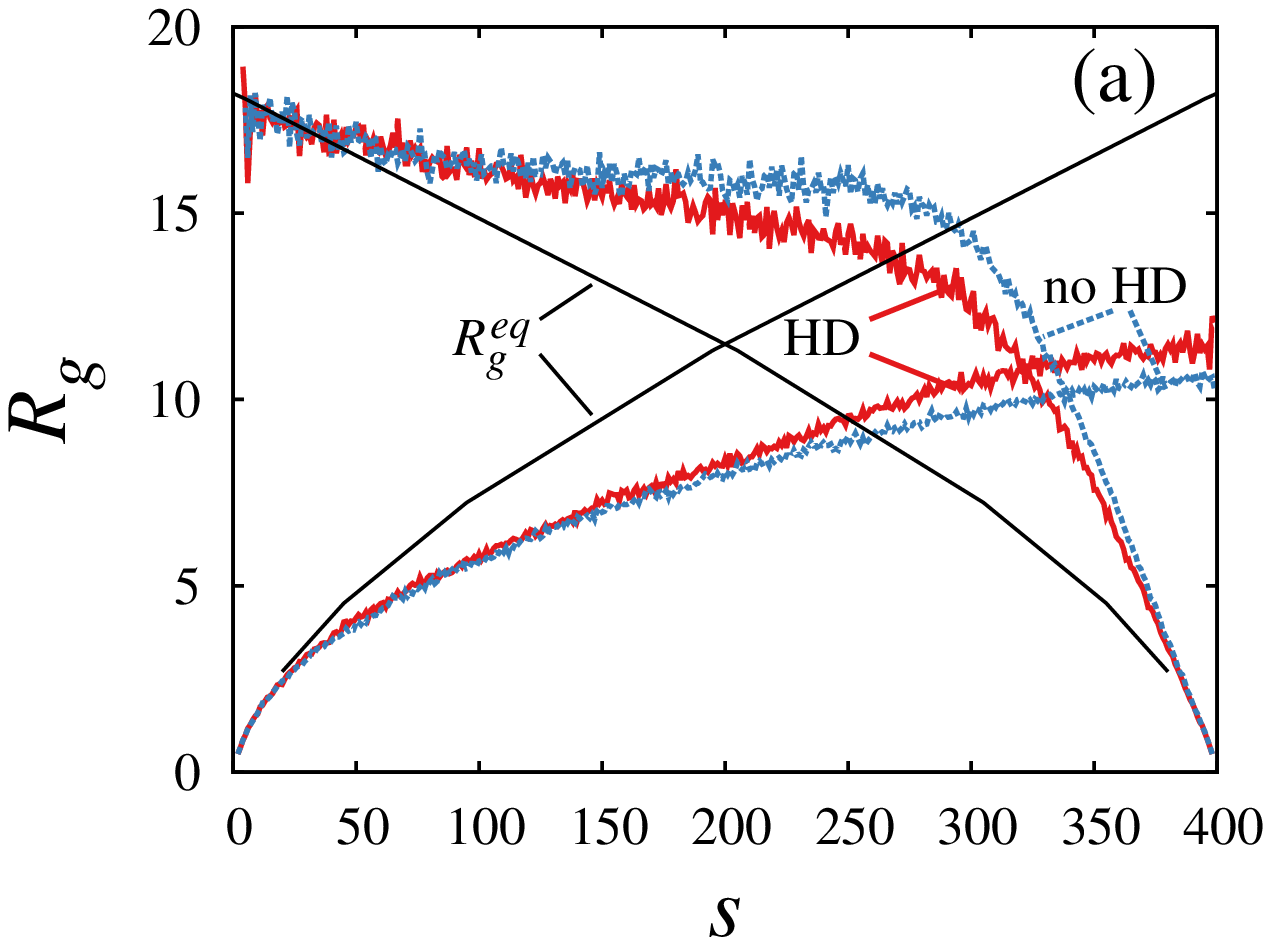}
\includegraphics[width=0.23\textwidth]{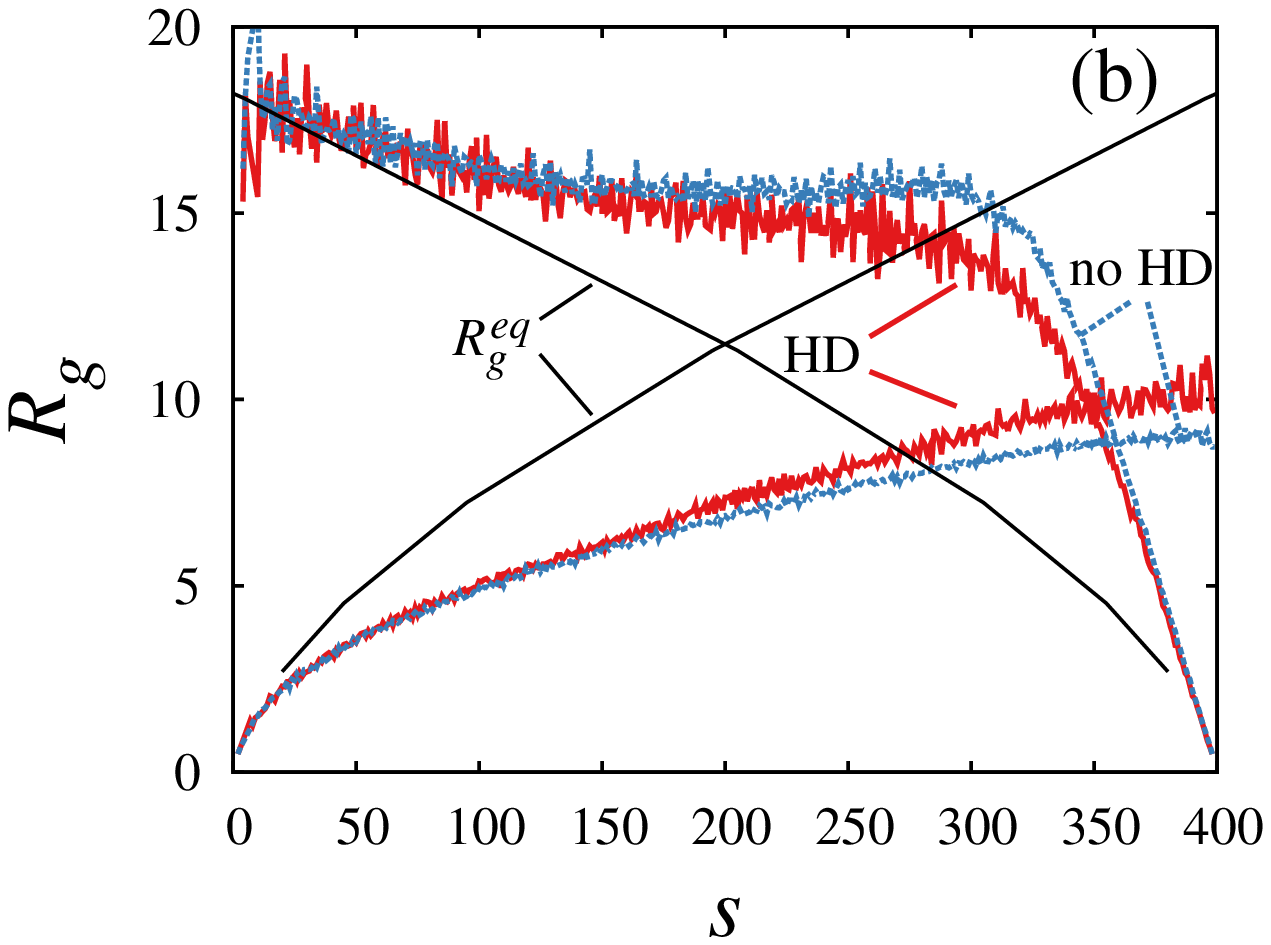}
\caption{\label{fig:rgcmp} Good solvent. Measured $R_g$ as a function of the translocation coordinate on the {\it trans} (plots increasing with $s$) and {\it cis} (plots decreasing as $s$ increases) sides. The equilibrium values for the radius of gyration $R_g^{eq}$ are are plotted with solid black lines. Hydrodynamics is included, $N = 400$. (a) $F = 1$. (b) $F = 3$.}
\end{figure}

Fig.~\ref{fig:rgcmp} shows $R_g$ measured separately on the {\it cis} and {\it trans} sides as a function of the number of translocated beads $s$ for the good solvent together with the plotted $R_g^{eq}(s)$, that is, $R_g$ for equilibrated conformations of $s$ monomers. (For the bad solvent $R_g(s) \approx R_g^{eq}(s)$ for all $s$ and are accordingly not shown.) In accordance with previous findings, $R_g$ deviates increasingly from the equilibrium value $R_g^{eq}$ on both sides in the course of the translocation, see {\it e.g.}~\cite{lehtola09}. The measured $R_g$ on both sides deviate more from $R_g^{eq}$ when hydrodynamics is not included. On the {\it cis} side $R_g > R_g^{eq}$  due to the straightening of the polymer, that is, tension propagation. On the {\it trans} side $R_g < R_g^{eq}$ due to crowding of monomers.

The crowding results from the polymer exiting the pore faster to the {\it trans} side than it relaxes to equilibrium. Since the deviation from equilibrium is slightly larger for polymers simulated without hydrodynamics, hydrodynamics speeds up the relaxation of the polymer to thermal equilibrium more than it speeds up translocation. For $F = 10$ the situation changes (not shown). Hydrodynamics speeds up translocation more effectively for larger $F$. The speed of relaxation to equilibrium does not change with $F$. Accordingly, the polymer segment on the {\it trans} side is driven further out of equilibrium for larger $F$, as seen in Fig.~\ref{fig:rgcmp}. For the very large $F = 10$ (not shown) the polymer conformation on the {\it trans} side is equally strongly compressed whether hydrodynamics is included or not.

\subsection{\label{sec:tension}Waiting times, tension, and friction}

\subsubsection{Good solvent}

It is well established that in good solvent polymer translocation dynamics is in practice determined by the tension propagating along the polymer chain on the {\it cis} side~\cite{PhysRevE.76.021803,lehtola09,PhysRevE.78.061803,PhysRevE.81.041808,rowghanian11:_force,PhysRevE.85.041801,PhysRevE.85.051803}. The mechanism was introduced in~\cite{PhysRevE.76.021803} and first extracted from simulations by registering the number of moving beads $n_d$~\cite{lehtola09}. Recently, we measured tension propagation more accurately and showed that in the absence of hydrodynamics a quasi-static model was sufficient to describe the process with great precision~\cite{suhonen14}. The method used in~\cite{lehtola09} was based on monomer velocity measurements to determine $n_d$. Due to back-flow effects one cannot obtain precise dynamics of tension propagation with this method when hydrodynamics is included. In the present study we investigate tension propagation when including hydrodynamic interactions using the more precise and direct method used in~\cite{suhonen14}. This way we obtain tension propagation with good precision also in the presence of hydrodynamics and can relate it to the better understood case of tension propagation without hydrodynamic interactions. 

\begin{figure}
\includegraphics[width=0.33\textwidth]{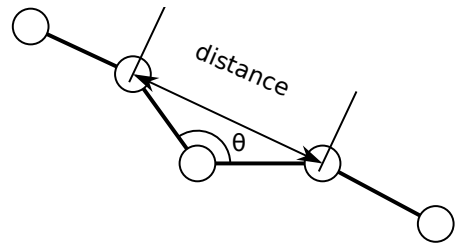}
\caption{\label{fig:drag-distance} A schematic five-bead polymer segment showing the measured drag distance over the three middle beads.}
\end{figure}

A more direct way to obtain tension propagation dynamics than measuring monomer velocity is to measure strain between all two consecutive beads in the polymer. The measured values of the bead-to-bead distances turned out to fluctuate strongly and hence require averaging over a vast number of measurements. For this reason, we measure the local straightening of a polymer over three consecutive beads as depicted in Fig.~\ref{fig:drag-distance}. We define the drag distance as $d_i = \left\| \beadpos_{i+1} - \beadpos_{i-1} \right\|$, which is directly proportional to the strain of a single bond. The average distance for polymers in equilibrium is $d_i \approx 2^\flory \ljsigma$. Obviously, for a completely straightened polymer segment $d_i = 2 \ljsigma$ ~\cite{suhonen14}.

\begin{figure}
\includegraphics[width=0.23\textwidth]{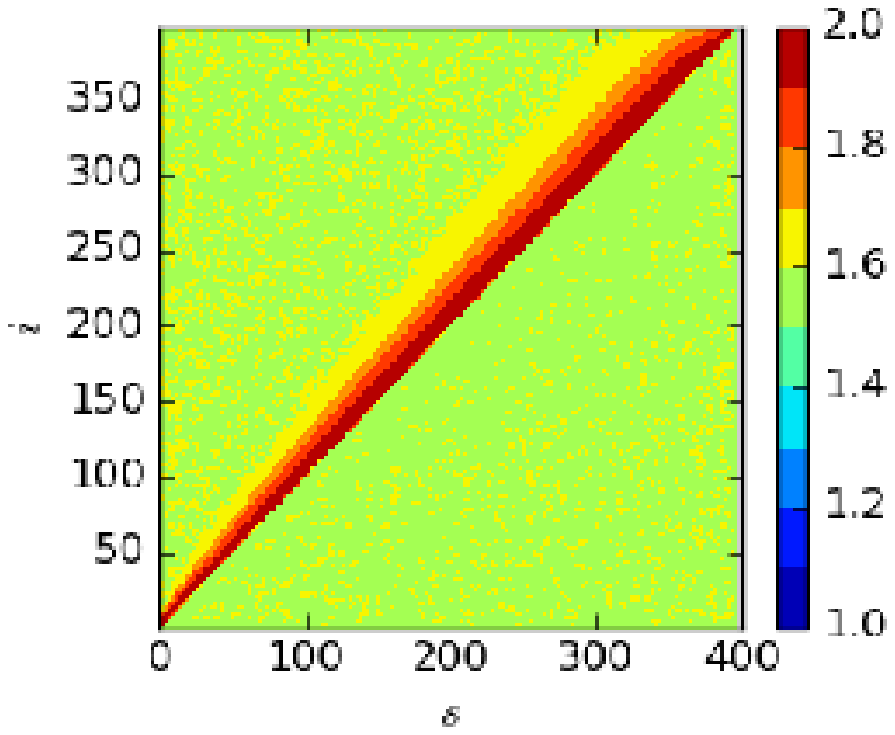}
\includegraphics[width=0.23\textwidth]{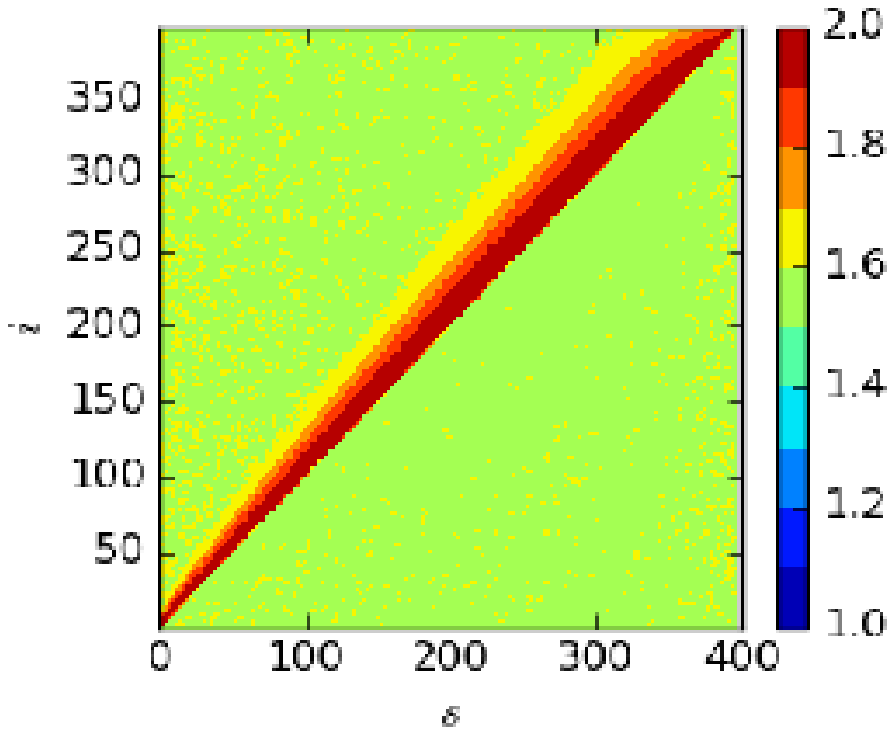}
\includegraphics[width=0.23\textwidth]{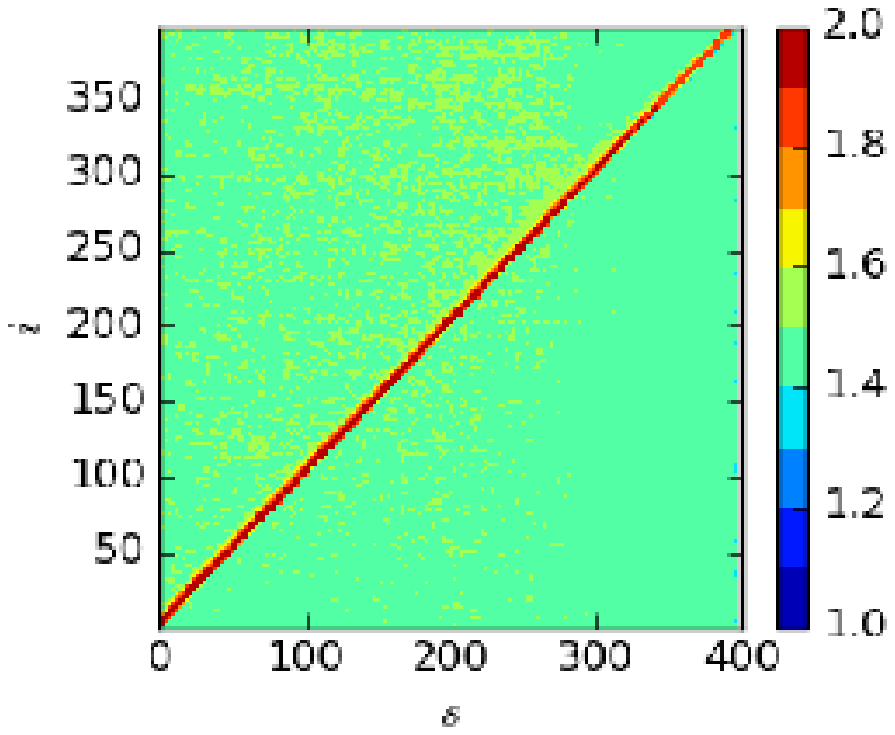}
\includegraphics[width=0.23\textwidth]{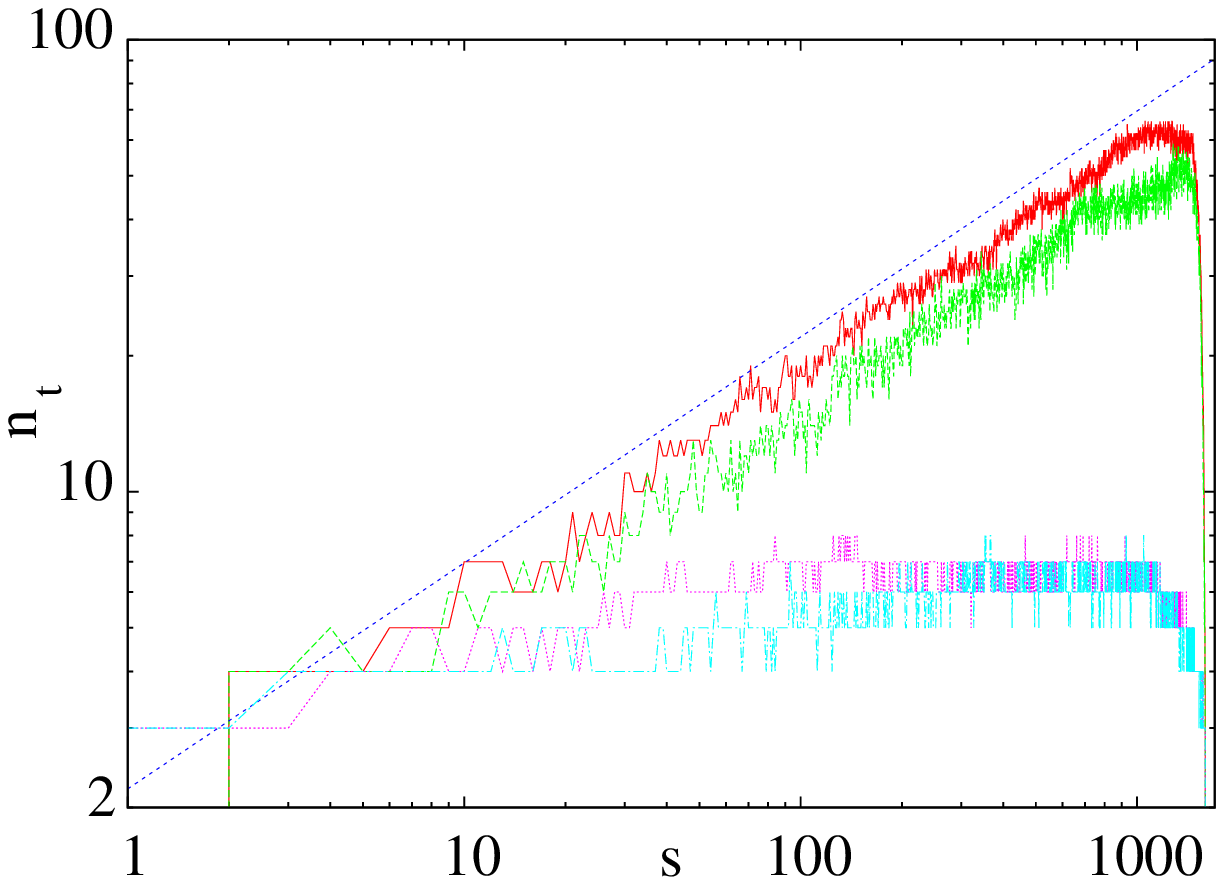}
\caption{\label{fig:tension} {\bf The first three figures}: The drag distances $d_i$ of each bead $i$ as a function of the translocation coordinate $s$. $F = 10$ and $N =  400$. Top left: for good solvent with hydrodynamics. Top right: for good solvent without hydrodynamics. Bottom left: for bad solvent with hydrodynamics. {\bf The last figure}: A logarithmic plot of the number of beads in the tensed segment $n_t$ on the {\it cis} side as a function of $s$. $F = 3$ and $N =  1600$. Curves from top to bottom: Good solvent without hydrodynamics, good solvent with hydrodynamics, bad solvent without hydrodynamics, and bad solvent with hydrodynamics. The dotted line shows the scaling $\sim s^{0.5}$.}
\end{figure}

In the three color charts of Fig.~\ref{fig:tension} the distance $d_i$ is depicted for each bead $i$ as a function of the reaction coordinate $s$. Colors (or shades of gray in print) show the tension around the bead $i$ of the polymer at the time when the bead $s$ is at the pore. On the diagonal $i = s$ and the bead $i$ resides at the pore where the tension is the greatest. Above the diagonal line $i > s$ and this region depicts the tension on the {\it cis} side. Below the diagonal $i < s$, which corresponds to the {\it trans} side, where the polymer is not tensed.
  
The color charts on the first row are for polymers in good solvent with and without hydrodynamics. The pore force $F = 10$. Tension is seen to propagate qualitatively in the same way in both cases. However, in the presence of hydrodynamics tension propagates slightly less effectively along the chain. This is understandable, since the solvent with hydrodynamic modes mediates the momenta of the moving beads farther from the pore along the chain. Consequently, a bead with a given label $n$ will be set in motion earlier (at a smaller $s$) in a solvent supporting hydrodynamic modes than in a Brownian heat bath. This results in smaller tension for a given $s$ when hydrodynamics is included. 

The first plot on the second row in Fig.~\ref{fig:tension} shows $d_i$ as a function of $s$ for polymers translocating in a bad solvent that supports hydrodynamics for $F = 10$. As can be seen tension does not propagate appreciably, the only discernible tension seen only on the diagonal, that is, at the pore entrance on the {\it cis} side. The plot for the case where polymer translocates in a Brownian heat bath with no hydrodynamics is similar (not shown).

We define the number of beads in the tensed segment $n_t$ on the {\it cis} side as the number of consequent beads starting from the bead $s$ at the pore  for which $d_i$ exceeds the equilibrium value, $d_i > 2^\nu \sigma \approx 1.5$. In other words, $n_t$ is the difference of the label of the last bead belonging to this segment and the current value of the translocation coordinate $s$. $n_t$ is the measure of the length of the tensed segment on the {\it cis} side. For improved resolution we use the threshold value $1.65$ for the bond length as a criterion for its end beads to belong to the tensed segment. The number of beads in drag $n_d$ that was measured in~\cite{lehtola09} differs from $n_t$, especially when hydrodynamics is included, as already pointed out. $n_d$ is not measured here but used to denote the actually moving polymer segment that determines the friction on the {\it cis} side. 

The last plot in Fig.~\ref{fig:tension} shows $n_t$ as a function of $s$ for $F = 3$.  The tension is seen to spread more effectively in the absence of hydrodynamic interactions even in the case of bad solvent where tension spreading is negligible. From this logarithmic plot it is also evident that the spreading tension gives the resultant scaling of the translocation time $\tau \sim N^\beta$ under good solvent conditions.

\begin{figure}
\includegraphics[width=0.23\textwidth]{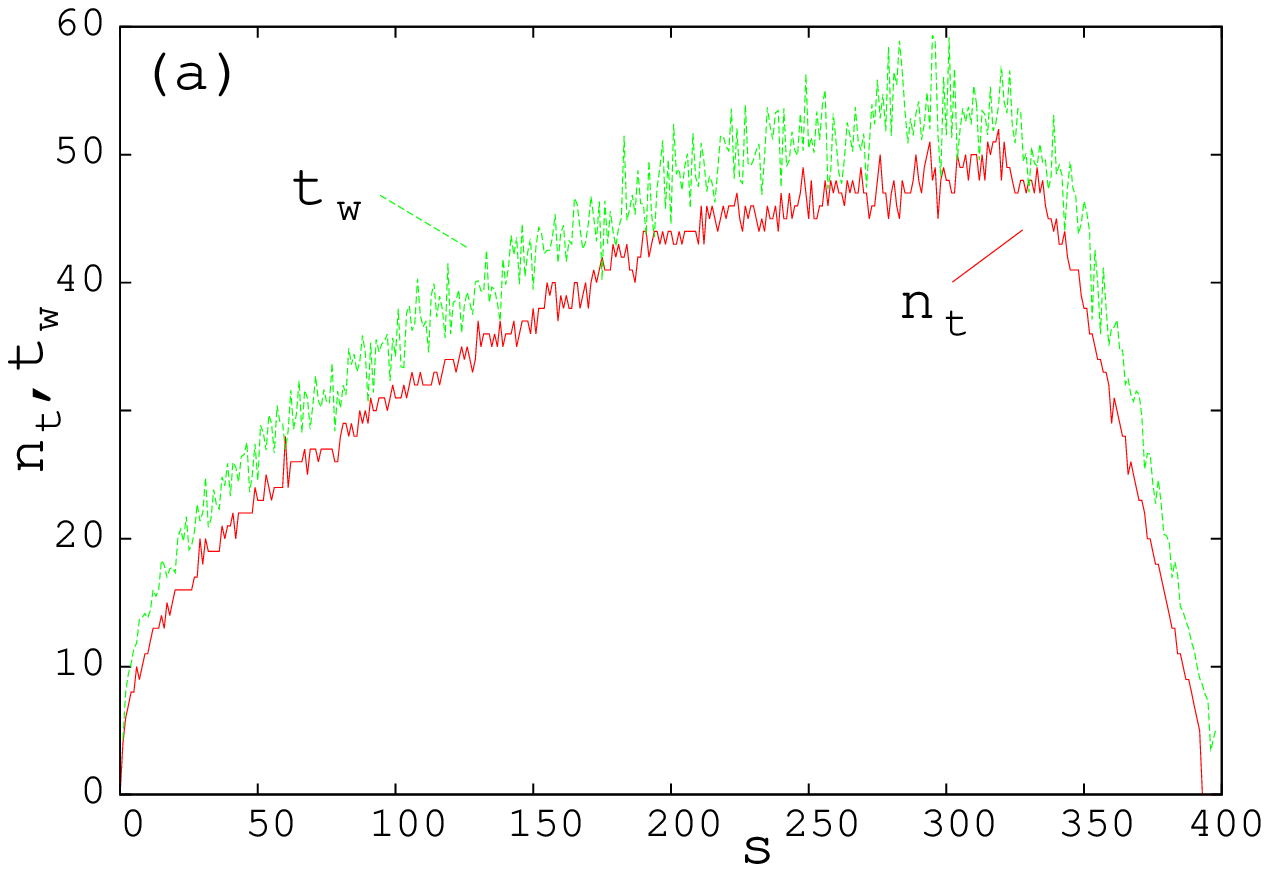}
\includegraphics[width=0.23\textwidth]{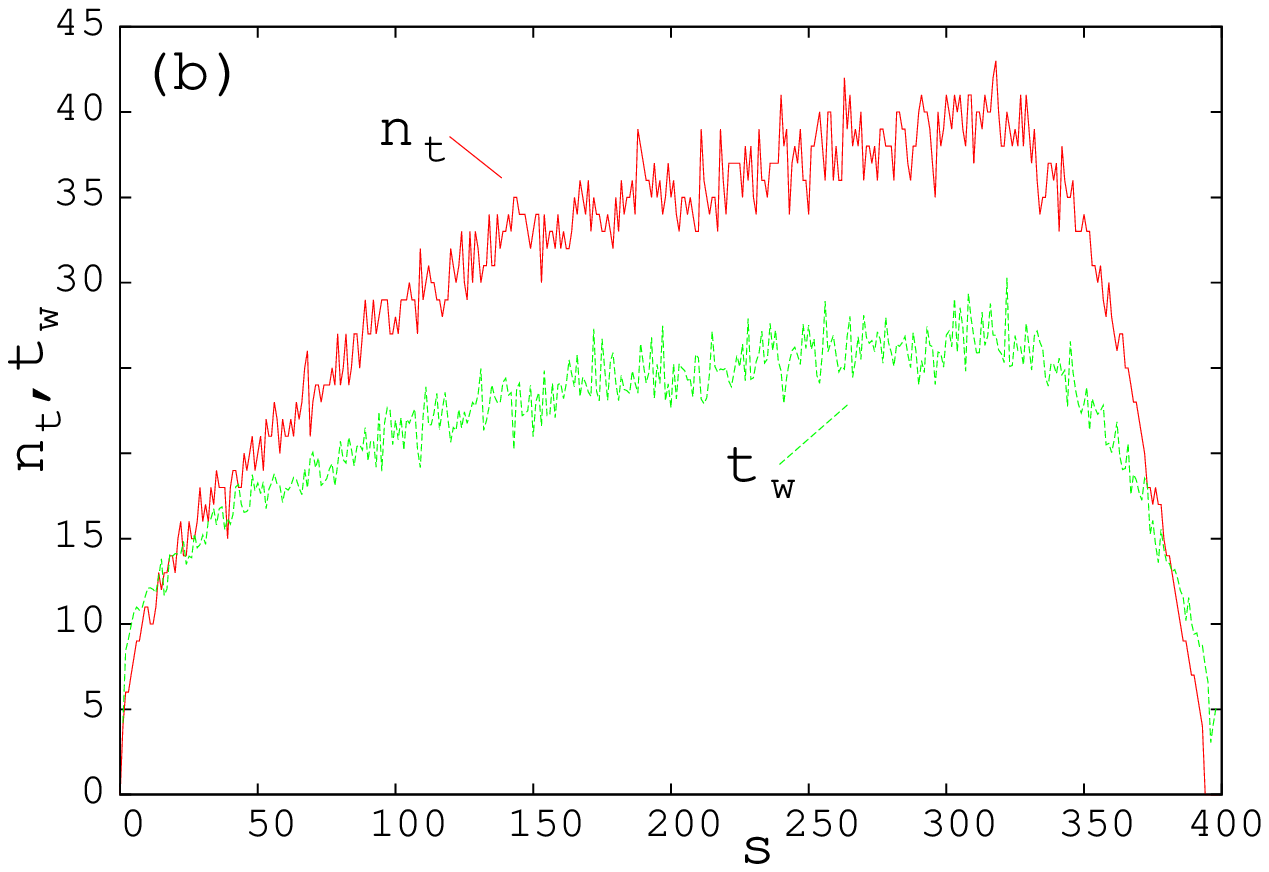}
\includegraphics[width=0.23\textwidth]{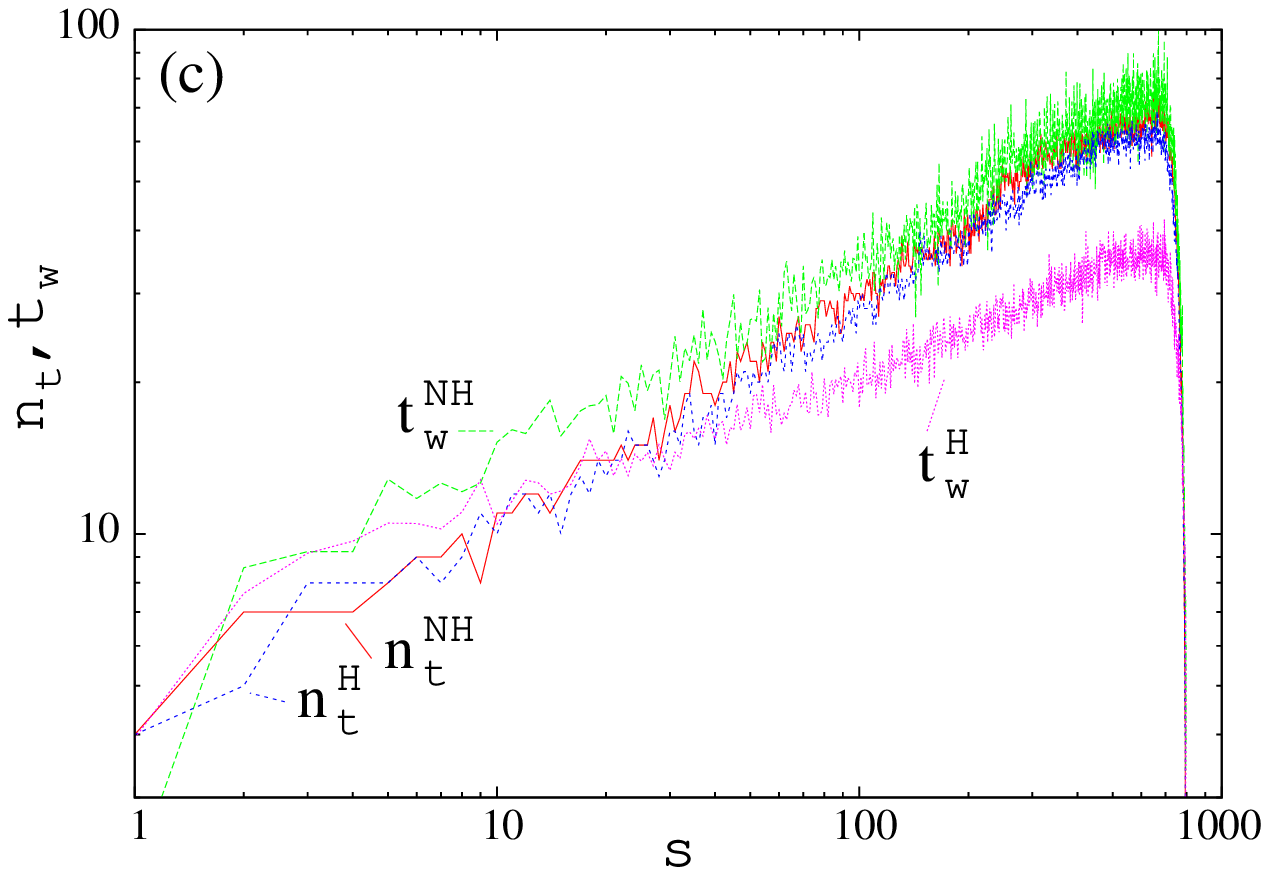}
\includegraphics[width=0.23\textwidth]{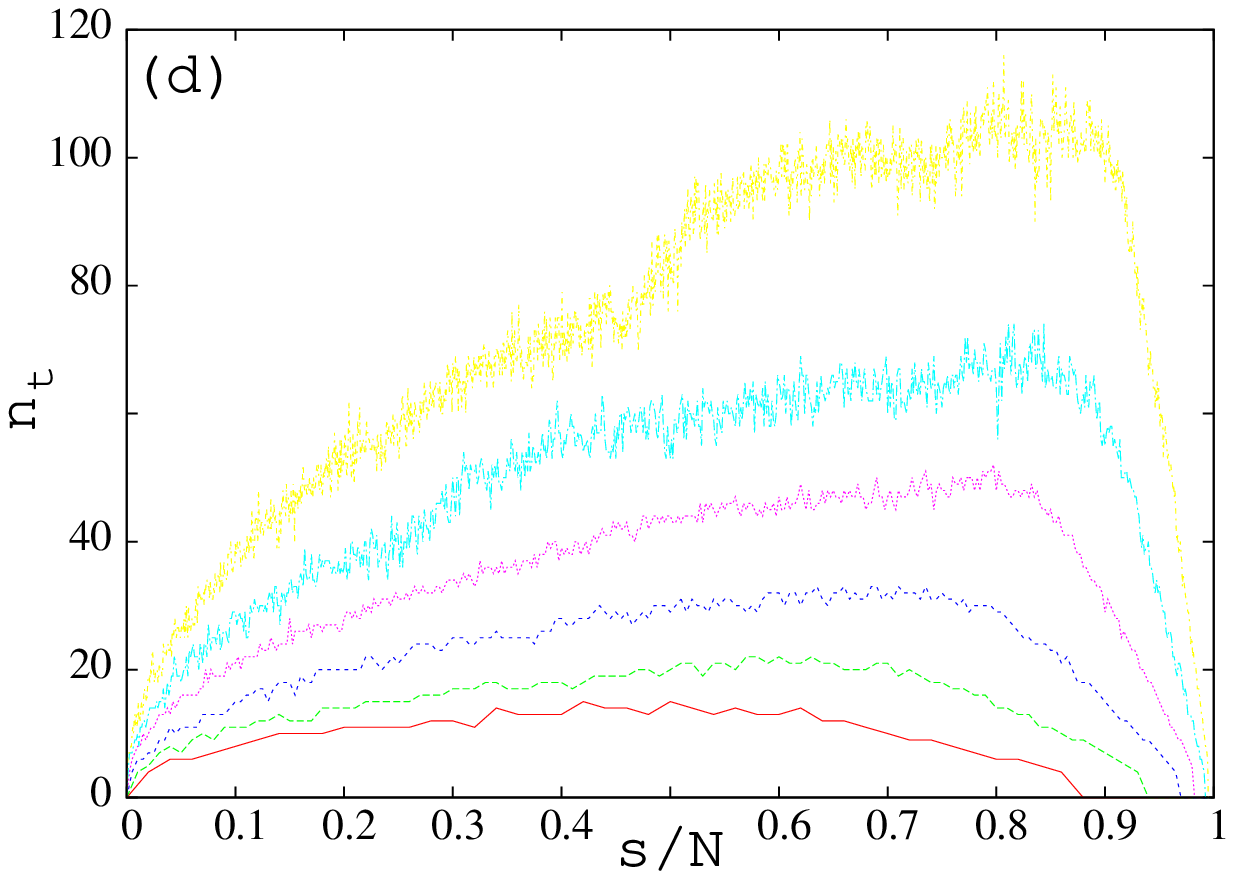}
\caption{\label{fig:wcplots} Comparison of the number of beads in the tensed segment $n_t$ and the waiting times $t_w$ as a function of the translocation coordinate $s$ in good solvent. $F = 3$. (a) Hydrodynamics not included. $N = 400$. (b) Hydrodynamics included. $N = 400$. (c) Logarithmic plots of $n_t(s)$ and $t_w(s)$ with (H) and without hydrodynamics (NH). $N = 800$. (d) $n_t$ as a function of the normalized reaction coordinate $s/N$ with no hydrodynamics, $F = 3$. From bottom to top $N = 50$, $100$, $200$, $400$, $800$, and $1600$.}
\end{figure}

The waiting time $t_w(s)$ is defined as the time required for the bead $s$ to reach the pore after the instant when the bead $s-1$ has entered the pore. In Figs.~\ref{fig:wcplots}~(a)-(c) the number of beads in the tensed segment $n_t(s)$ and waiting times $t_w(s)$ are compared for the translocation in good solvent driven by $F = 3$. In the absence of hydrodynamic interactions, see Fig.~\ref{fig:wcplots}~(a), $n_t(s)$ depends on $s$ in almost exactly the same way as $t_w$, that is, $n_t(s) \propto t_w(s)$, which is in agreement with the results from our previous study using Langevin dynamics. In this study we showed that a quasi-static model, where the tension spreading is described only geometrically and inertial and stochastic components are ignored describes the process fairly accurately~\cite{suhonen14}.

The relation $t_w \propto n_t$ in a good solvent without hydrodynamics is a consequence of the driven translocation taking place strongly out of equilibrium~\cite{lehtola09}. Hence, the process can be largely described by how the frictional force $\eta$ changes on the {\it cis} side. In the quasi-static model~\cite{suhonen14} the beads from $s$ to $s + n_t$ are considered to be the only non-stationary beads affecting the translocation dynamics, the other beads being at rest. The driving force must balance the friction experienced by $n_d$ beads moving with velocity $v$ towards the pore $F = \eta = n_d \gamma v$, or $v^{-1} \propto n_d$. Hence, when the bead $s-1$ is at the pore, the time required for the next bead $s$ to reach the pore is given by $t_w(s) \approx \ljsigma / \vel(s) \propto n_d(s) = n_t(s)$. The last equality holds in the absence of hydrodynamics. The translocation time $\tau = \int_0^{N-1} t_w(s) ds$. Accordingly, if $n_t(s) \sim s^\xi$ in the tension propagation phase, then disregarding the final retraction of the polymer tail $\tau \sim N^\beta$, where $\beta = \xi+1$. From the last plot in Fig.~\ref{fig:tension} we can read $\beta \approx 1.5$.

Hydrodynamic interactions break the relation $t_w \propto n_t$, as seen in Fig.~\ref{fig:wcplots}~(b). The friction increasing with $n_t$ is still a determining factor for the waiting time profile $t_w(s)$. However, due to the backflow beads start moving before tension reaches them. The backflow of the moving beads sets beads in motion farther from the pore than where the tension has propagated. This leads to smaller $n_t$ in the presence of  hydrodynamics. Consequently, $n_t < n_d$ when hydrodynamics is included. Also, in contrast to the translocation dynamics in the Brownian heat bath, the friction is not directly proportional to $n_d$ when hydrodynamics is included but determined by the hydrodynamic radius of the moving polymer segment.

In Fig.~\ref{fig:wcplots}~(b) $n_t$ and $t_w$ are compared with and without hydrodynamics. $t_w(s)$ is seen to initially increase as rapidly in the presence and absence of hydrodynamics. It takes a while for the hydrodynamic modes to fully develop after the first polymer beads are set in motion. Before this the translocating polymers immersed in solvents with and without hydrodynamics experience identical friction. After this setting-in time for hydrodynamics the friction for the translocating polymer in the solvent with hydrodynamics is much smaller. Consequently, the $t_w(s)$ profile grows more weakly with $s$ than the $n_t(s)$ profile. This is also clearly seen in the logarithmic plots of $n_t(s)$ and $t_w(s)$ in Fig.~\ref{fig:wcplots}~(c). $n_t(s)$ scale identically for the cases with and without hydrodynamics. In the absence of hydrodynamics $t_w(s)$ is closely aligned with $n_t(s)$, whereas $t_w(s)$ scales with $s$ with a clearly smaller exponent than $n_t(s)$ when hydrodynamics is included.

During the translocation the waiting time $t_w$ increases to a maximum before dropping rapidly due to the final retraction of the remaining polymer segment on the {\it cis} side. The ratio for the waiting times in the absence and presence of hydrodynamics $R = t^{noHD}_w(s)/t^{HD}_w(s)$ reaches a maximum $R_{max}$ at this same point. For $N \le 800$, $R_{max}$ does not change with $F$ but increases with $N$. We obtain $R_{max} = 1.2$, $1.3$, $1.5$, $1.7$, $2$, and $2$ for $N = 25$, $50$, $100$, $200$, $400$, and $800$. This characteristics is explained by tension not having a sufficient time to evolve before retraction starts for short polymers, as seen from Fig.~\ref{fig:wcplots}~(d) showing $n_t(s/N)$ for different $N$ in  the absence of hydrodynamics. Consequently, the tension profiles in the absence and presence of hydrodynamics differ less for short polymers.

Only for $N=1600$ values of $R_{{\rm max}}$ differ slightly for different $F$; namely $R_{{\rm max}} = 3$ for $F = 1$ and $R_{{\rm max}} = 2.3$ for $F = 3$ and $10$. As seen in Fig.~\ref{fig:wcplots}~(d), translocation of the polymer of length $N = 1600$ is occasionally stalled already for $F = 3$. This stalling affects not only $R_{{\rm max}}$ but also $\beta$, for $F =10$.

\begin{figure}
\includegraphics[width=0.23\textwidth]{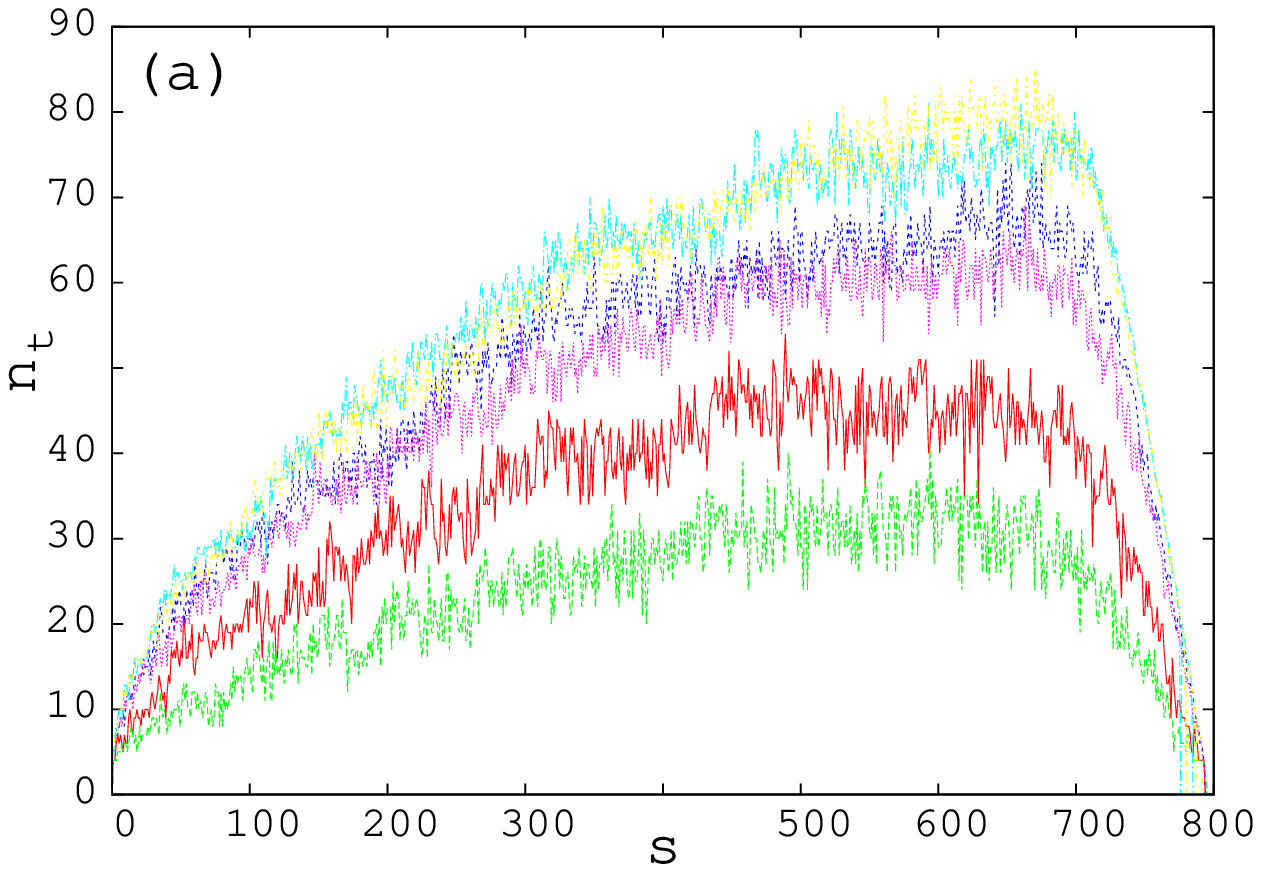}
\includegraphics[width=0.23\textwidth]{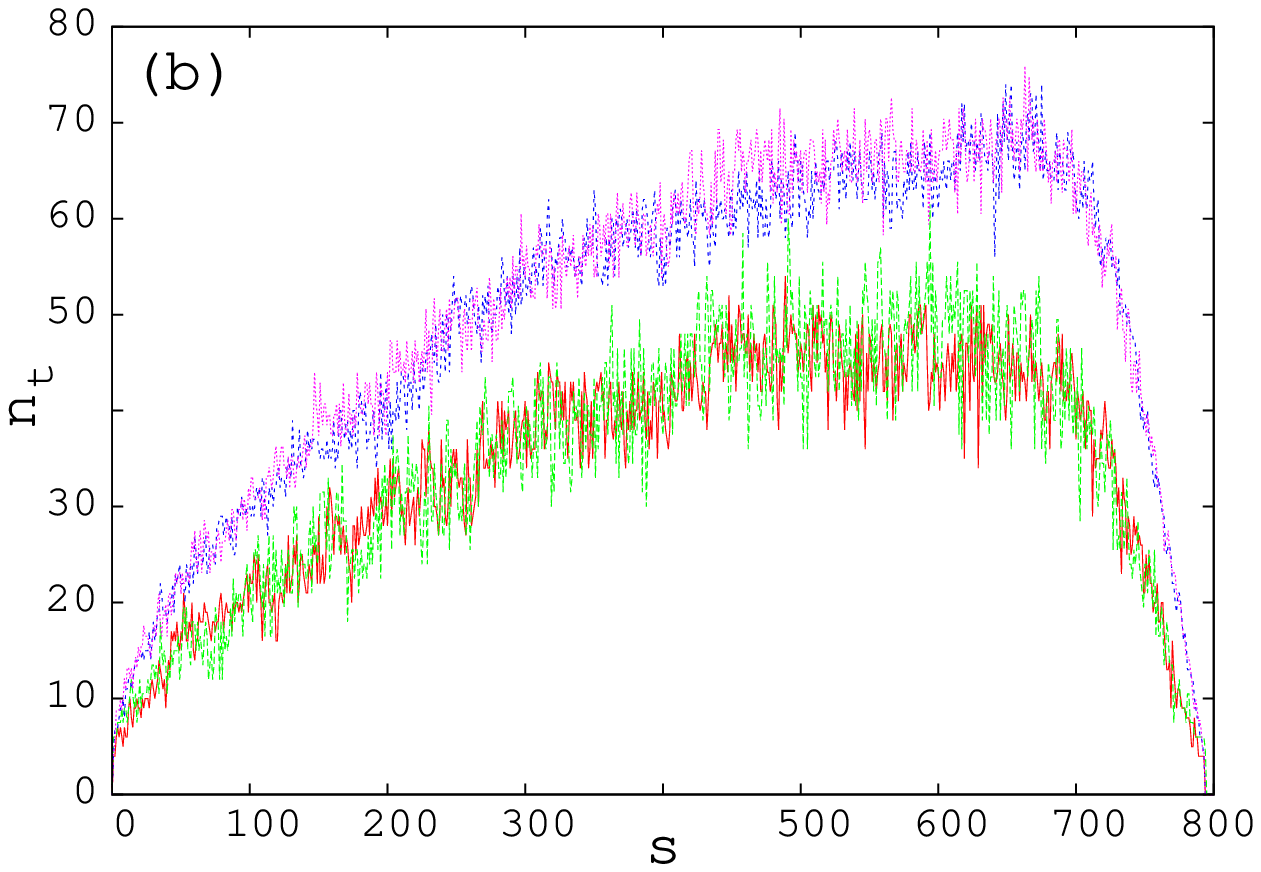}
\caption{Good solvent, $N = 800$. (a) The length of the tensed segment on the {\it cis} side during translocation $n_t(s)$. Curves from bottom to top: (i) $F=1$, with hydrodynamics (HD), (ii) $F=1$,  no HD, (iii) $F=3$ with HD, (iv) $F=3$ no HD, (v \& vi) $F=10$ with and without HD. (b) Tension profiles $n_t(s)$ with ad without HD. $n_t(s)$ with HD are scaled, $A n_t(s)$, such that perfect alignment with no-HD-profiles are obtained. $A = 1.5$ for $F=1$ (lower curves) and $A = 1.1$ and $F=3$ (upper curves).}
\label{fig:tensionF}
\end{figure}

In Fig.~\ref{fig:tensionF} we look more closely at the lengths of the tensed segments. $n_t$ increases with pore force $F$, see Fig.~\ref{fig:tensionF}~(a). In other words, tension propagates faster for larger driving pore force. It is noteworthy that the tension propagates farther for larger $F$ in spite of the fact that also the polymer translocates faster and so spends less time on the {\it cis} side. Hence, increasing $F$ speeds up tension propagation more than polymer translocation.

As seen in Fig.~\ref{fig:tensionF}~(a) the increase of $n_t$ with increasing $F$ gets weaker for larger $F$. When a bead at the end of the tensed segment is being pulled by a moderate force, the next bonded bead has time to  move in the viscous heat bath before the bond is fully stretched, or in our measurement the two-bond drag distance $d_i= 2\sigma$ (see Fig.~\ref{fig:drag-distance}). When this force increases the bond gets more and more extended before the next bead starts moving due to the diminished effect of stochasticity and possibly increased effect of inertia. For a sufficiently large force the bond gets fully extended before the next bead moves. $n_t$ cannot increase beyond the full extent given by the quasi-static model so by increasing $F$ we will approach this model where beads that are set in motion from their initial positions start immediately moving at the velocity of the pulling bead~\cite{suhonen14}.

This change of $n_t$ with $F$ is a deviation from the quasi-static model and contributes to a dependence found in a number of studies, namely that the translocation time is not strictly inversely proportional to the driving force, but instead $\tau \sim F^{-\alpha}$, where $\alpha < 1$. A detailed study of this effect is beyond the scope of the present paper and will be conducted in a forthcoming paper~\cite{pauli_new}. 

For $F = 1$ and $F = 3$ tension propagates more strongly in the absence of hydrodynamic interactions. The difference in tension propagation for the cases with and without hydrodynamics is largest for small $F$ and decreases when increasing $F$ until it disappears for $F = 10$, where tension propagates so fast that it overrides the effect of the hydrodynamic backflow, whose effect is seen for $F = 1$ and $3$.

Although the magnitudes of $n_t(s)$ with and without hydrodynamics are different the shapes are identical, which can be seen in Fig.~\ref{fig:tensionF}~(b), where the tension profiles $n_t(s)$ for $F = 1$ and $3$ with hydrodynamics are scaled to make them align with the corresponding $n_t(s)$ without hydrodynamics. In other words, tension propagates along the polymer chain on the {\it cis} side similarly during the translocation for the cases with and without hydrodynamics. Only the magnitudes of $n_t(s)$ are smaller when hydrodynamics is involved for $F < 10$. The length of the tensed segment without hydrodynamics is by the factor $A = 1.5$ larger than with hydrodynamics for $F = 1$. For $F = 3$ $A = 1.1$. For $F = 10$ tension profiles for translocations with and without hydrodynamics are identical ($A \approx 1.0$).


In order to have some idea of the relative importance of the two factors contributing to the speed-up due to hydrodynamics, namely the backflow and the reduction of friction, we performed simulations where a polymer in free solvent (no walls, periodic boundaries) is pulled at the end by constant force $f_{\rm drag}$. These simulations were started from straight polymer conformations that are relatively close to the quasi-static conformations that the polymers assume after being dragged for a sufficiently long time. We checked that similar conformations (and terminal velocities) were obtained by starting from equilibrated conformations. The velocity of the polymer was measured after it had reached the terminal value, as we have previously done for sedimenting polymers~\cite{piili_sedimentation}. The measurement was done for polymers of different lengths and for different $f_{\rm drag}$ (10 runs per parameter pair).

\begin{figure}
\includegraphics[width=.49\textwidth]{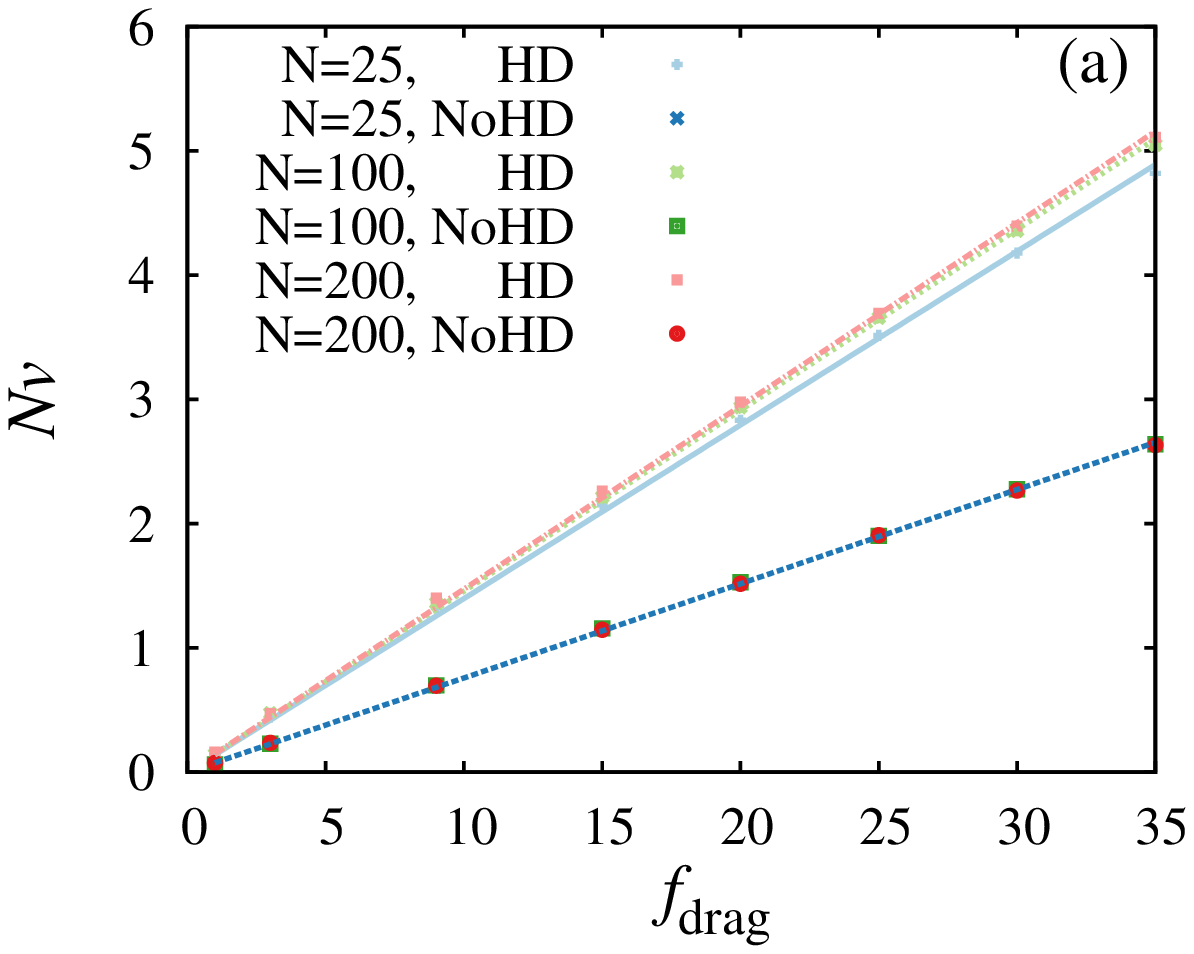}
\includegraphics[width=.49\textwidth]{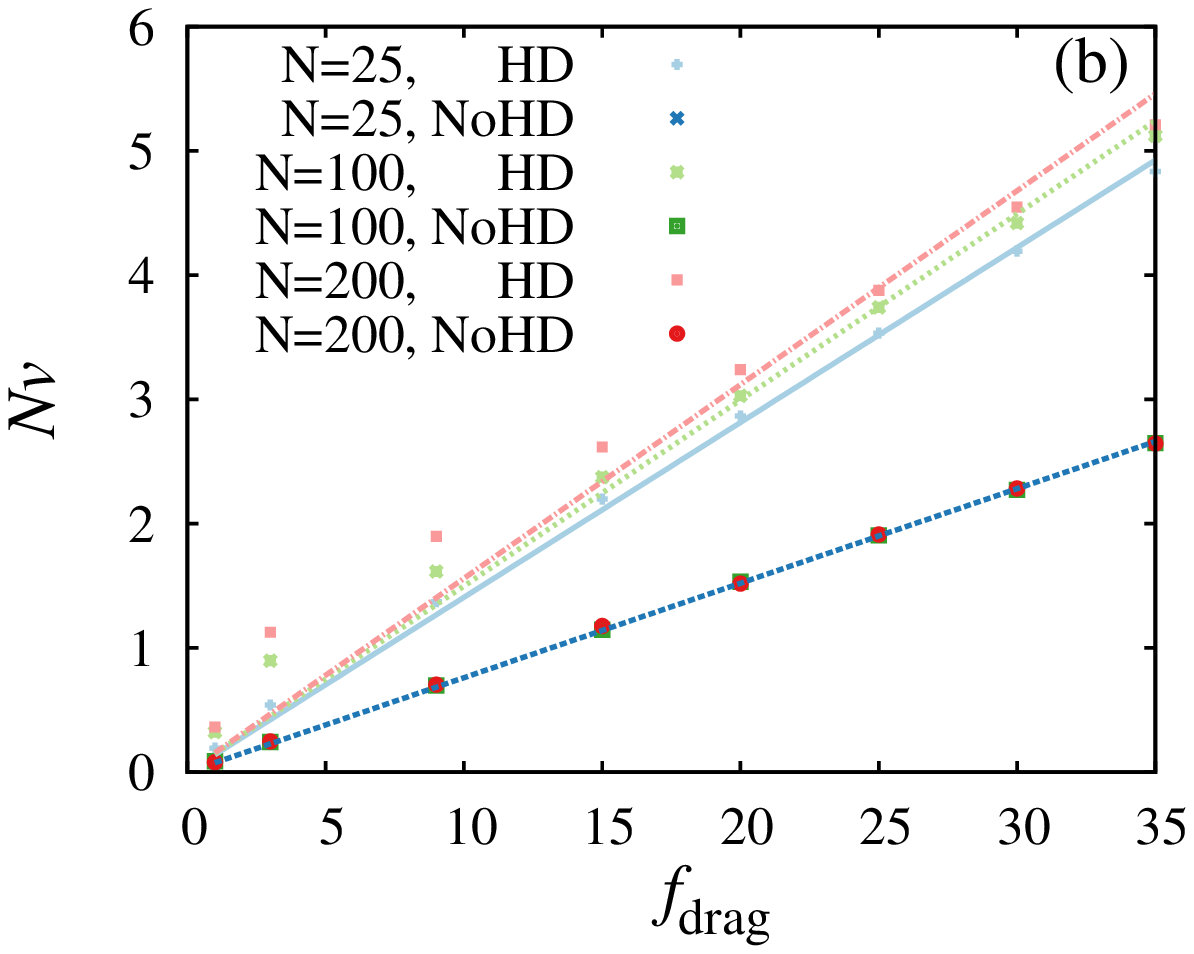}
\caption{\label{fig:dragPlots} Terminal velocity times polymer length $N v$ as a function of dragging force $f_{\rm drag}$ for polymers of different lengths. Upper plots: hydrodynamics included. Lower plots: hydrodynamics not included. The dashed lines are linear fits to the data for $N = 25$. (a) Good solvent. (b) Bad solvent. In the case of bad solvent with hydrodynamics the least-squared fitted lines are plotted to show the deviation from the relation $Nv \sim f_{\rm drag}$ (Eq.(~\ref{eq:dragforce}).)}
\end{figure}

The measured components of the terminal velocities parallel to $f_{\rm drag}$ follow quite accurately the relation
\begin{align}\label{eq:dragforce}
v = C_{\rm(HD/noHD)} \frac{f_{\rm drag}}{N},
\end{align}
where the constants $C_{\rm{HD}} \approx 0.15$ and $C_{\rm noHD} = 0.076$  are for the cases with and without hydrodynamics, respectively. In Fig.~\ref{fig:dragPlots}~(a) the terminal velocities multiplied by the polymer lengths $N v$ are plotted as a function of $f_{\rm drag}$. (This is done instead of plotting $v$ vs $f_{\rm drag}/N$ to more clearly show the deviations from the relation given by Eq.~(\ref{eq:dragforce}).) The values for the coefficients are given in Table~\ref{tab:coeff}. Hydrodynamics is seen to speed up the motion of the dragged polymers of lengths $N \in [25, 400]$ by a factor $C_{\rm HD}/C_{\rm noHD} \approx 1.9 $. Since the effect of backflow is insignificant in the case of a fully extended polymer being pulled at the end, this is the ratio of the frictions in the absence and presence of hydrodynamics. From Fig.~\ref{fig:wcplots}~(d) and comparing $C_{\rm HD}/C_{\rm noHD}$ to the maximum ratio of waiting times $R_{{\rm max}}$ above hydrodynamic modes can be estimated to have time to develop fully before retraction for polymers of length $N \ge 400$. The backflow plays a role at the initial and intermediate stages of translocation, but before the retraction at the end of the tension propagation the driven polymer translocation for a realistic pore bias has reached a state where hydrodynamics speeds up the motion solely due to the reduced friction.

\begin{table}
\caption{The constants $C$ of Eq.~(\ref{eq:dragforce}) for polymers of
different length $N$ with and without hydrodynamics (HD) and their ratios. Good (GS) and bad solvent (BS).}
\begin{tabular}{ccccccc}
$N$ & $C_{\rm HD}^{\rm GS}$ & $C_{\rm noHD}^{\rm GS}$ & $C_{\rm HD}^{\rm GS}/C_{\rm noHD}^{\rm GS}$ & $C_{\rm HD}^{\rm BS}$ & $C_{\rm noHD}^{\rm BS}$ & $C_{\rm HD}^{\rm GS}/C_{\rm noHD}^{\rm BS}$ \\
1 &  0.0735 & 0.0731 & 1.01 & 0.0735 & 0.0731 & 1.01\\
25 & 0.140 & 0.0758 & 1.84 & 0.141 & 0.0760 & 1.85\\
100 & 0.146 & 0.0759 & 1.92 & 0.150 & 0.0760 & 1.97\\
200 & 0.147 & 0.0757 & 1.95 & 0.156 & 0.0762 & 2.05\\
400 & 0.148 & 0.0755 & 1.96 & - & - & -\\
\end{tabular}
\label{tab:coeff}
\end{table}

\subsubsection{Bad solvent}
\label{sec:bs}

\begin{figure}
\includegraphics[width=.49\textwidth]{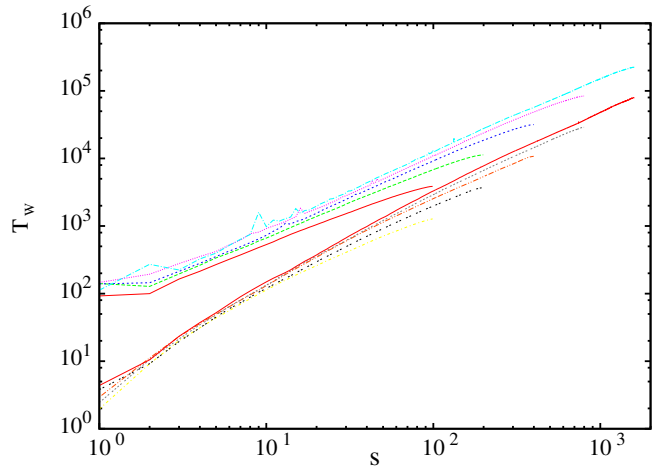}
\caption{\label{fig:cumwt} Translocation in bad solvent, no hydrodynamics. Logarithmic plot of cumulative waiting times $T_w(s)$ for $F =1$ (upper curves) and $3$ (lower curves). Polymer lengths $N = 100$, $200$, $400$, $800$, and $1600$.}
\end{figure}

In the bad solvent tension does not appreciably propagate along the polymer chain on the {\it cis} side, in contrast to the translocation in good solvent, see Figs.~\ref{fig:transrg} and \ref{fig:tension}. Accordingly, there is no scaling of $\tau$ with $N$ due to tension propagation. This is clearly seen in Fig.~\ref{fig:cumwt} showing cumulative waiting times, $T_w(s) = \int_0^s t_w(s') ds'$ for $F = 1$ and $3$ and $N \in [100, 1600]$. ($T_w(s)$ is the average time it takes for the bead $s$ to enter the pore.)

Right from the beginning of translocation the waiting times are seen to be larger for longer chains. This reflects the correlated motion of the polymer beads in the globular (and entangled) conformation in the bad solvent. Unlike in the good solvent the correlation length in the bad solvent extends over the whole globular polymer segment on the {\it cis} side (see the second row in Fig.~\ref{fig:snapshots}). Hence, a single moving monomer having to push other monomers out of its path experiences a friction that increases with the number of monomers on the {\it cis} side. The waiting times increasing for any $s$ when increasing $N$ is in stark contrast with waiting times of translocations in good solvent, where for polymers of different $N$ they are aligned and follow $t_w \sim s^\gamma$  (up to the point where the final retraction from the {\it cis} side starts). Consequently, in the bad solvent the waiting time profile does not result in the scaling of $\tau$ with $N$ like in the good solvent.

\begin{figure}
\includegraphics[width=.29\textwidth]{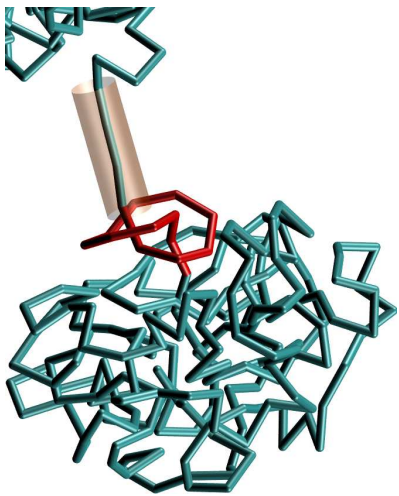}
\caption{\label{fig:knot} A snapshot of a simulation with bad solvent at the reaction coordinate value $s=1400$. A trefoil knot (emphasized in red) is seen at the pore entrance. $N=1600$ and $F = 3$. For improved visibility the polymer is depicted with a tube of diameter $0.2$ that is smaller than the diameter of $1$ implied by the Lennard-Jones interaction range.}
\end{figure}

Also in the bad solvent simulations where polymers were pulled at the end by a constant force $f_{\rm drag}$ were made. Fig.~\ref{fig:dragPlots}~(b) shows the terminal velocity $v$ multiplied by the polymer length $N$ as a function of the pulling force $f_{\rm drag}$ in the case of bad solvent for $N = 25$, $100$, and $200$. As was the case in the good solvent, the polymers start from a straight conformation. For strong $f_{\rm drag}$ the conformations remain fairly straight but for weak $f_{\rm drag}$ polymers have a globular portion. In the absence of hydrodynamics individual monomers experience a similar friction regardless of the length of the pulled polymer. This is seen as a collapse of $Nv$-$f_{\rm drag}$ curves for polymers of different $N$.

In the presence of hydrodynamics monomers belonging to longer polymers obtain higher velocities than those belonging to shorter ones. Also, polymers pulled by a constant force obtain higher velocities in a bad than in a good solvent. Both these features are due to the backflow assisting a polymer with a prominent globular part. The effect of the backflow is not significant for the straight polymers in the good solvent. The coefficients in Eq.~(\ref{eq:dragforce}) for the bad-solvent case are shown in the three rightmost columns of Table~\ref{tab:coeff}.
 
In spite of hydrodynamics assisting collective motion of a globular polymer segment in the pulling experiment, it has a negligible effect on driven translocation in the bad solvent. In the case of translocation individual monomers that are pulled toward the pore are surrounded by immobile monomers, so that long-ranged hydrodynamic modes in the direction of the motion do not form. Accordingly, although hydrodynamics speeds up collective center-of mass motion of entire polymer conformations even more effectively in the bad than in the good solvent, it has a negligible effect on the driven translocation in the bad solvent.

It is also noteworthy that for polymers of realistic lengths knots slow down translocation. In our simulations stalling effects in individual translocations due to knots are seen for polymers of lengths $N = 800$ and $1600$. Fig.~\ref{fig:knot} shows a snapshot of a simulated translocation where a trefoil knot is at the pore entrance.


\section{\label{sec:conclusion}Summary}

We have studied driven polymer translocation in the good and the bad solvent in the presence and absence of hydrodynamic modes using stochastic rotation dynamics coupled with molecular dynamics. By measuring the radii of gyrations $R_g$ on both sides of the pore we found that during the translocation polymers were not driven out of equilibrium in the bad solvent, in contrast to the translocation in the good solvent. The mechanisms driving the polymer out of equilibrium in the good solvent are well established, namely tension propagation on the {\it cis} and monomer crowding on the {\it trans} side. In the bad solvent polymer translocated through the pore practically from one equilibrium conformation to another. Here, translocations driven by pore force of large magnitudes were seen to be occasionally stalled by knots present already in the initial random conformations.

In the good solvent the {\it trans} side polymer conformations deviated from equilibrium conformations less when hydrodynamics was included. This means that hydrodynamics speeds up polymer's relaxation toward equilibrium even more than it speeds up its translocation. Also on the {\it cis} side $R_g$ of the polymer conformations deviate less from the equilibrium values when hydrodynamics is involved. This is a manifestation of the backflow setting monomers in motion before the tension propagating from the pore has reached them. Consequently, the polymer is less tensed, or less extended, on the {\it cis} side in the presence of hydrodynamics.

Tension measurements showed that the length of the tensed segment increases with the same exponent in the presence and absence of hydrodynamics. In other words, tension propagates qualitatively in the identical manner whether hydrodynamics is included or not. Only the length of the tensed segment at all stages is smaller due to backflow when hydrodynamics is included. For a constant pore force the inclusion of hydrodynamics reduces the length of the tensed segment by a constant factor. This is quite remarkable, since it means that the difference {\it e.g.} in the scaling relations come only from the difference in friction that the polymer experiences depending on whether hydrodynamics is included or not. The polymer conformations for both case are identical. This information is useful for inclusion of hydrodynamics in the present theories of driven polymer translocation. 

In spite of the tension propagating identically with the the number of translocated monomers in the presence and absence of hydrodynamics the waiting times in the two cases differ. A smaller exponent for the scaling of the translocation time with the polymer length is obtained in the presence of hydrodynamics as is well known. In the beginning of the translocation waiting time profile follows the tension propagation also when hydrodynamics is included. This is due to the set-in time required for the hydrodynamic modes to fully develop. After this initial stage the friction is smaller when hydrodynamics is included and, consequently, the waiting time that is to a good precision proportional to the friction scales with the number of translocated monomers with a smaller exponent in the presence of hydrodynamics.

In the bad solvent the tension does not propagate appreciably on the {\it cis} side. Accordingly, the translocation time does not show a clear scaling with the polymer length in the bad solvent. The waiting times at the initial stages of the translocations are longer for longer polymers, which is due to the increased friction that the moving monomers experience by the relatively immobile monomers in their path within the globular conformations. However, for very long polymers linear dependence of the translocation time on the polymer length seems to be approached.

By measuring terminal velocities for polymers in free solvent pulled at the end, we could determine the speed-up due to hydrodynamics in the good solvent to result almost entirely from the reduced friction and to a lesser extent from the backflow. Surprisingly, hydrodynamics was seen to speed up the pulled polymer in the bad solvent more than in the good solvent. In addition, the motion of longer polymers was sped up more than the motion of short polymers. This is explained by the globular polymer conformation moving 'as a whole' in the bad solvent, that is, by collective center-of-mass motion of the monomers, whereas in the translocation individual monomers move within the globular conformation. The backflow is much stronger in the motion of a globular than a stretched polymer. (Compare a droplet-shaped vehicle to a long truck.) Hence, in the presence of hydrodynamics the pulled polymers move faster in the bad than in the good solvent and the longer polymers with a sufficiently large globular part move faster than short polymers.

To recap, in the good solvent hydrodynamics does not essentially change the polymer conformations during the translocation, only scales down by a constant factor the extended, or tensed, polymer segment on the {\it cis} side. Hence, all differences come from the different friction experienced by the essentially same polymer conformations in the presence and absence of hydrodynamics. In the bad solvent  there is no tension propagation nor the accompanying scaling of the translocation time $\tau$ with the polymer length $N$ in the driven translocation. Hydrodynamics that in the good solvent is known to speed up translocation and decrease the scaling exponent $\beta$ has only a negligible effect in the bad solvent. In the biological context the polymer translocation takes place in solutions abundant with biological organelles, which can make the solvent quality effectively bad. Hence, the scaling characteristics obtained for the generic translocation in the good solvent may not be completely valid for the {\it in vivo} translocation processes.

\begin{acknowledgments}
Mr.~Pauli Suhonen is acknowledged for his contribution in the useful discussions, especially those concerning tension propagation. The computational resources of CSC-IT Centre for Science, Finland, and Aalto Science-IT project are acknowledged. The work of Joonas Piili is supported by The Emil Aaltonen Foundation and The Finnish Foundation For Technology Promotion. 
\end{acknowledgments}

\bibliography{paper}

\end{document}